\documentclass[11pt,letterpaper]{article}
\usepackage[dvipdfmx]{graphicx}
\usepackage{comment}
\usepackage{amsmath}
\usepackage{amsfonts}
\usepackage{amsthm} 
\usepackage{chicago}
\usepackage{caption}
\usepackage{here}
\usepackage{color}

\makeatletter
\newcommand{\subsubsubsection}{\@startsection{paragraph}{4}{\z@}%
  {1.0\Cvs \@plus.5\Cdp \@minus.2\Cdp}%
  {.1\Cvs \@plus.3\Cdp}%
  {\reset@font\sffamily\normalsize}
}
\makeatother
\def\cite{\def\@citeseppen{-1000}%
    \def\@cite##1##2{##1\if@tempswa , ##2)\else{)}\fi}%
    \def\citeauthoryear##1##2##3{##1 (##3}\@citedata}

\newtheorem{lem}{Lemma}[section]

\newtheorem{prop}{Proposition}[section]

\usepackage{algorithm}
\usepackage{algorithmicx}
\usepackage{algpseudocode}

\newcounter{nombre} 
\renewcommand{\thenombre}{\arabic{nombre}} 
\newcommand{\remark}[1][]{\refstepcounter{nombre}#1{\it Remark \thenombre}. }

\allowdisplaybreaks[0]

\title{Particle rolling MCMC with double-block sampling}
\author{Naoki Awaya\thanks{Graduate School of Economics, The University of Tokyo, Tokyo, Japan.}\ \ and \ \ Yasuhiro Omori\thanks{Faculty of Economics, The University of Tokyo, Japan. E-mail: omori@e.u-tokyo.ac.jp}}

\date{\today}

\begin{document}
\maketitle
\thispagestyle{empty}
\begin{abstract}
An efficient particle Markov chain Monte Carlo methodology is proposed for the rolling-window estimation of state space models. The particles are updated to approximate the long sequence of posterior distributions as we move the estimation window.  To overcome the well-known weight degeneracy problem that causes the poor approximation, we introduce a practical double-block sampler with the conditional sequential Monte Carlo update where we choose one lineage from multiple candidates for the set of current state variables. Our proposed sampler is justified in the augmented space through theoretical discussions. In the illustrative examples, it is shown to be successful to accurately estimate the posterior distributions of the model parameters.
\end{abstract} 

\newpage
\setcounter{page}{1}
\section{Introduction}
State space models have been popular and widely used in the analysis of economic and financial time series. These models are flexible and capture the dynamics of the complex economic structure. However, several structural changes have been noted in long-term economic series. If the precise time of a structural change is known, we could divide the sample period into two periods, before and after the structural change. However, this time point is usually unknown, and the change may occur gradually from one state to another.  Although there are various statistical models for the structural change in the literature, the rolling-window estimation is the simple and common way to reflect the recent change in the forecasting without delay where we fix the number of observations to estimate model parameters and update the dataset to improve the forecasting performance.

In non-linear or non-Gaussian state space models, the likelihood is often not obtained analytically, and the maximum likelihood estimation is difficult to implement. The Markov Chain Monte Carlo (MCMC) method is a popular and powerful technique used to estimate model parameters and state variables by generating random samples from the posterior distribution given a set of observed data for various complex state space models. However, for rolling estimation, simply applying the MCMC method would be too time-consuming given the need to estimate a long sequence of posterior distributions.

To overcome this difficulty, we take an alternative approach based on the sequential Monte Carlo (SMC) sampler discussed in \shortciteN{DelMoral2006}.
This is effective because, in the rolling-window estimation, we can utilize the weighted samples from one posterior distribution to approximate the next posterior distribution instead of reiterating the same MCMC algorithm with the slightly different dataset. The particles consist of realized values of state variables and static parameters, which are updated when including a new observation and excluding the old observation. 
As we shall show in the illustrative examples of Section 4, a simple rolling-window sampler that is derived in a straightforward manner from the previous literature leads to the severe weight degeneracy problem, suggesting that the updating step should be constructed carefully. To fix this problem, we adopt the idea of block sampling (e.g. \shortciteN{Doucet2006}, \shortciteN{Polson2008}), in which state variables at multiple time points are updated simultaneously when learning new information. It is highly efficient in the sense that it substantially increases the effective sample size. Based on this idea, we propose the novel sampling method, called the double-block sampler, where we sample a block of state variables when both including and excluding the information.

However, unless the time series model has a relatively simple form, finding an appropriate proposal distribution for these update steps may be difficult. Hence, instead of generating only one candidate from the proposal distribution, we generate multiple candidates and choose one of them using the conditional SMC of the particle MCMC (\shortciteN{AndrieuDoucetHolenstein(10)}). This nested structure is similar to that of SMC$^2$ (\shortciteN{Chopin2013}, \shortciteN{Fulop2013}) and nested SMC \shortcite{naesseth2015nested}, but our proposed algorithm differs in that it is derived from the particle Gibbs instead of the particle MH (Metropolis-Hastings) algorithm.
As a special case of our new method, our proposed double-block sampler can be used to implement the ordinary sequential analysis by keeping all past observations. It contrasts with SMC$^2$ in that it originates from different types of the particle MCMC algorithms.

The remainder of the paper is organized as follows. In Section 2, we introduce the simple rolling-window sampler for state space models and point out that such a sampler derived from the conventional filtering algorithm causes the serious weight degeneracy phenomenon. Section 3 introduces a double-block sampler to  overcome this difficulty.  Section 4 provides illustrative examples and, in Section 5, theoretical justifications of the proposed method are provided. Section 6 concludes the paper.
\section{Particle rolling MCMC in general state space models}
\subsection{Rolling-window estimation in general state space model}
Consider the state space model which consists of a measurement equation, a state equation with an observation vector $y_t$, and an unobserved state vector $x_t$ given a static parameter vector $\theta$. For the prior distribution of $\theta$, we let $p(\theta)$ denote its prior probability density function.
Further define $x_{s:t}\equiv (x_s,x_{s+1},\ldots,x_t)$ and $y_{s:t}\equiv (y_s,y_{s+1},\ldots,y_t)$.
We assume that the distribution of $y_t$ given $(y_{1:t-1}, x_{1:t}, \theta$) depends exclusively on $x_{t}$ and $\theta$ and that the distribution of $x_t$ given ($x_{1:t-1}$, $\theta$) depends only on $x_{t-1}$ and $\theta$. The corresponding probability density functions are noted as follows:
\begin{eqnarray}
p(y_t \mid x_{1:t}, y_{1:t-1}, \theta) &=& p(y_t \mid x_t,\theta) \equiv g_\theta(y_t \mid x_t), \quad t=1,\ldots,n,\\
p(x_t \mid x_{1:t-1},\theta) &=& p(x_t \mid x_{t-1},\theta) \equiv f_\theta(x_t \mid x_{t-1}), \quad t=2,\ldots,n,
\end{eqnarray}
where $p(x_1 \mid \theta) \equiv \mu_{\theta}(x_1)$ denotes a known density function of the stationary distribution given $\theta$.

We also incorporate the correlation between $y_t$ and $x_{t+1}$, which is conditional on $x_t$ since we consider such an example, the realized stochastic volatility (RSV) model,  for the financial time series (see e.g. a seminal work by \shortciteN{Takahashi2009}) in our illustrative example. It is a stochastic volatility model with an additional measurement equation for the realized volatility. Let $y_t=(y_{1,t},y_{2,t})'$ where $y_{1,t}$ and $y_{2,t}$ denote the daily log return and the logarithm of the realized volatility (variance) at time $t$. Let $x_t$ denote the latent log volatility which is assumed to follow the stationary AR(1) process. The RSV model is defined as follows:
\begin{eqnarray}
\label{eq:rsv-1}
y_{1,t} &=& \exp(x_t / 2)\epsilon_t,\ \epsilon_t \sim \mathcal{N}(0,1),\ t = 1,\dots,T \\
\label{eq:rsv-2}
y_{2,t} &=& x_t + \xi + u_t,\ u_t \sim \mathcal{N}(0, \sigma_u^2),\ t = 1,\dots,T \\
\label{eq:rsv-3}
x_{t+1} &=&  \mu + \phi(x_t - \mu) +  \eta_t,\ \eta_t \sim \mathcal{N}(0,\sigma^2_\eta),\ t = 1,\dots,T,\\
\label{eq:rsv-4}
x_1 &=& \mu + \frac{1}{\sqrt{1-\phi^2}}\eta_0,\ \eta_0 \sim \mathcal{N}(0,\sigma^2_\eta), \quad |\phi|<1,
\end{eqnarray}
where
\begin{eqnarray}
\label{eq:rsv-5}
\left(
   \begin{array}{c}
     \epsilon_t \\
     u_t \\
     \eta_t 
   \end{array}
\right)
\sim
\mathcal{N} 
\left(
\left[
   \begin{array}{c}
     0 \\
     0 \\
     0 
   \end{array}
\right],
\left[
   \begin{array}{ccc}
     1 & 0 & \rho \sigma_\eta \\
     0 & \sigma^2_u & 0 \\
     \rho \sigma_\eta & 0 & \sigma^2_\eta  
   \end{array}
\right]
\right),
\end{eqnarray}
$\mathcal{N}(\mu,\Sigma)$ denotes a normal distribution with mean $\mu$ and covariance matrix $\Sigma$, 
and $\theta=(\mu,\phi,\sigma^2_\eta,\xi,\sigma^2_u,\rho)'$ is the static parameter vector. 
\noindent
The correlation $\rho$ between $\epsilon_t$ and $\eta_t$ is introduced to express the leverage effect. The effect is often negative in empirical studies, which implies that the decrease in the today's log return is followed by the increase in the log volatility on the next day (e.g. \shortciteN{Omori2007}).
In this case, we express the dependence of $y_t$ on $x_{t+1}$ (or $x_{t+1}$ on $y_t$) as follows.
\begin{eqnarray}
p(y_t \mid x_{1:t+1},y_{1:t-1},\theta) &=& p(y_t \mid x_t,x_{t+1},\theta)\equiv g_\theta(y_t \mid x_t, x_{t+1}), \quad t=1,\ldots,n,\\
p(x_{t+1} \mid x_{1:t},y_{1:t}, \theta) &=& p(x_{t+1} \mid x_{t}, y_t,\theta)\equiv f_\theta(x_{t+1} \mid x_{t},y_t),
\quad t=1,\ldots,n-1.
\end{eqnarray}

In the rolling-window estimation of time series, the number of observations (or the window size) in the sample period is fixed and is set equal to, e.g., $L+1$. We estimate the posterior distribution of $\theta$ and $x_{s:t}$ given the observations $y_{s:t}$ with $t=s+L$ for $s=1,2\ldots$, and its probability density function is given by
\begin{eqnarray}
\pi(x_{s:t},\theta \mid y_{s:t}) &\propto & p(\theta)\mu_\theta(x_{s})g_\theta(y_{s} \mid x_{s})\left\{ \prod^t_{j={s+1}} f_\theta(x_j \mid x_{j-1},y_{j-1})g_\theta(y_j \mid x_j) \right\}, 
\end{eqnarray}
or, equivalently,
\begin{eqnarray}
\pi(x_{s:t},\theta \mid y_{s:t}) 
&\propto &   p(\theta)\mu_\theta(x_{s})\left\{\prod^t_{j=s+1} f_\theta(x_j \mid x_{j-1})g_\theta(y_{j-1} \mid x_{j-1},x_{j}) \right\}g_\theta(y_t \mid x_t).
\end{eqnarray}
\subsection{Simple rolling-window sampler}
\label{sec:A general framework}
We first describe a simple rolling-window sampler that is derived in a straightforward manner from the previous literature.
The estimation procedure consists of two steps, each of which can be described in the framework of the SMC sampler in \shortciteN{DelMoral2006} as follows. In Step 1, suppose we have samples from the old target density $\pi(x_{s-1:t-1},\theta\mid y_{s-1:t-1})$ with importance weight $W_{[s-1,t-1]}$ at time $t-1$  where the subscript $[s-1,t-1]$ implies that the weight is based on observations $y_{s-1:t-1}$. After we include an observation $y_t$,  our new target density is  $\pi(x_{s-1:t},\theta \mid y_{s-1:t})$. Using the proposal kernel $K((x_{s-1:t-1},\theta),(x_{s-1:t},\theta))$, we update the weight
\begin{eqnarray}
\label{eq:weight_update}
W_{[s-1,t]} = 
\frac{\pi(x_{s-1:t}, \theta \mid y_{s-1:t})L((x_{s-1:t}, \theta),(x_{s-1:t-1}, \theta))}{\pi(x_{s-1:t-1}, \theta \mid y_{s-1:t-1})K((x_{s-1:t-1}, \theta),(x_{s-1:t}, \theta))}
\times W_{[s-1,t-1]},
\end{eqnarray}
where $L((x_{s-1:t}, \theta),(x_{s-1:t-1}, \theta))$ is the artificial backward Markov kernel with $L\equiv 1$ in this step. 
The first factor on the right hand side of Equation (\ref{eq:weight_update}) is the incremental weight to adjust that of the previous step.

In Step 2, we have samples from the old target density $\pi(x_{s-1:t}, \theta \mid y_{s-1:t})$ with the importance weight $W_{[s-1,t]}$ from Step 1. After we exclude the observation $y_{s-1}$, our new target density is $\pi(x_{s-1:t}, \theta \mid y_{s:t})$. Using the backward kernel $L((x_{s:t}, \theta),(x_{s-1:t}, \theta))=\pi(x_{s-1:t}, \theta \mid y_{s:t})/\pi(x_{s:t}, \theta \mid y_{s:t})=\pi(x_{s-1}\mid x_{s:t}, y_{s:t}, \theta)$, we update the weight
\begin{eqnarray}
W_{[s,t]} = 
\frac{\pi(x_{s:t}, \theta \mid y_{s:t})L((x_{s:t}, \theta),(x_{s-1:t}, \theta))}{\pi(x_{s-1:t}, \theta \mid y_{s-1:t})K((x_{s-1:t}, \theta),(x_{s-1:t}, \theta))}
\times W_{[s-1,t]},
\end{eqnarray}
where $K((x_{s-1:t}, \theta),(x_{s-1:t}, \theta))$ is the artificial proposal kernel with $K\equiv 1$, and discard $x_{s-1}$. 
Additionally, we can refresh all the particles with the MCMC when we observe the weight degeneracy, which is also 
 regarded as an importance sampling step in the SMC sampler with the unnormalized weight equal to one. We note that one can also update particles using the particle Gibbs sampler (\shortciteN{AndrieuDoucetHolenstein(10)}). Details are given below.\\

\noindent
{\bf Step 1.}  
Assume that, at time $t-1$, we have a collection of particles $(x_{s-1:t-1}^n,\theta^n)$ with the importance weight $W_{[s-1,t-1]}^n$, ($n=1,\ldots,N$) which is a discrete approximation of $\pi(x_{s-1:t-1}, \theta \mid y_{s-1:t-1})$.
We include a new observation $y_t$ in the information set and aim to sample from $\pi(x_{s-1:t}, \theta \mid y_{s-1:t})$. Given the current sample $(x_{s-1:t-1}, \theta)$ from $\pi(x_{s-1:t-1}, \theta \mid y_{s-1:t-1})$, we propose a candidate $x_t$ using some proposal density $q_{t,\theta}(x_t \mid x_{t-1},y_{t})$. Since the incremental weight is 
\begin{eqnarray*}
\frac{\pi(x_{s-1:t}, \theta \mid y_{s-1:t})}{\pi(x_{s-1:t-1}, \theta \mid y_{s-1:t-1})q_{t,\theta}(x_t \mid x_{t-1},y_{t})}
&=&
\frac{p(x_{t}, y_t  \mid x_{s-1:t-1}, y_{s-1:t-1},\theta)}{q_{t,\theta}(x_t \mid x_{t-1},y_{t}) p(y_t\mid y_{s-1:t-1})}
\\
&\propto&
\frac{f_{\theta}(x_t \mid x_{t-1}, y_{t-1})g_{\theta}(y_t \mid x_t)}{q_{t,\theta}(x_t \mid x_{t-1},y_{t})},
\end{eqnarray*}
we generate $x^n_t \sim q_{t,\theta^n}(x^n_{t} \mid x^n_{t-1},y_{t})$ and compute the importance weight
\begin{eqnarray}
W_{[s-1,t]}^n 
&\propto&  \frac{f_{\theta^n}(x_t^n \mid x_{t-1}^n, y_{t-1})g_{\theta^n}(y_t \mid x_t^n)}{q_{t,\theta^n}(x_t^n \mid x_{t-1}^n,y_{t})}\times
W_{[s-1,t-1]}^n.
\label{incremental weight 1}
\end{eqnarray}

\noindent
Finally, we compute some degeneracy criteria such as the effective sample size (ESS), 
\begin{eqnarray}
\rm{ESS}_{[s-1:t]} \equiv \left[\sum^N_{n=1}\left\{W^n_{[s-1,t]}\right\}^2\right]^{-1},
\end{eqnarray}
and the particles are resampled if $\mathrm{ESS} <cN$ (e.g. $c = 0.5$). \\
%
%

\noindent
{\bf Step 2}. We exclude the old observation $y_{s-1}$ from the information set, and aim to sample from $\pi(x_{s:t},\theta \mid y_{s:t})$ where the backward kernel is $\pi(x_{s-1:t},\theta \mid y_{s:t})/\pi(x_{s:t},\theta \mid y_{s:t})=\pi(x_{s-1}\mid x_{s:t},y_{s:t}, \theta)=p(x_{s-1} \mid  x_s, \theta)$ given by
\begin{align}
\label{eq:initial_density}
    p(x_{s-1} \mid x_s, \theta)
    &=
    \frac{\mu_\theta(x_{s-1}) f_\theta(x_s \mid x_{s-1} )}{\mu_\theta(x_s)}.
\end{align}
Since the current sample is from $\pi(x_{s-1:t}, \theta \mid y_{s-1:t})$ where
\begin{eqnarray*}
\pi(x_{s-1:t}, \theta\mid y_{s-1:t})&=&
\frac{p(x_{s-1:t},y_{s-1},\theta \mid y_{s:t})}{p(y_{s-1} \mid y_{s:t})}
=\frac{\pi(x_{s-1:t}, \theta \mid y_{s:t})p(y_{s-1}\mid x_{s-1:t},y_{s:t},\theta)}{p(y_{s-1}\mid y_{s:t})},
\end{eqnarray*}
the (unnormalized) incremental weight is
\begin{eqnarray*}
\frac{\pi(x_{s:t}, \theta \mid y_{s:t})p(x_{s-1} \mid  x_s, \theta)}{\pi(x_{s-1:t}, \theta \mid y_{s-1:t})}
&=&
\frac{p(y_{s-1} \mid y_{s:t})}{p(y_{s-1} \mid  x_{s-1:t},y_{s:t},\theta)}
\propto
g_{\theta}(y_{s-1} \mid x_{s-1},x_{s})^{-1},
\label{incremental weight 2}
\end{eqnarray*}
Thus, we update the importance weight
\begin{eqnarray}
\label{eq:weight-step2}
W_{[s,t]}^n &\propto&  g_{\theta^n}(y_{s-1} \mid x_{s-1}^n,x_{s}^n)^{-1} \times W_{[s-1,t]}^n,
\end{eqnarray}
and discard $x_{s-1}^n$.
If some degeneracy criteria are fulfilled, resample all the particles by implementing the MCMC algorithm as in Step 1.  The above procedure is summarized in Algorithm 1. \\

%
\begin{table}[H]
\centering
{\bf Algorithm 1: Simple rolling-window sampler}
\\
\footnotesize
\hspace{0.4cm}
\hrulefill
\vspace{-2mm}
\\
\begin{enumerate}
\item[]
Let  $(x_{s-1:t-1}^n,\theta^n)$ denote the sample from $\pi(x_{s-1:t-1}, \theta \mid y_{s-1:t-1})$ with the weight $W_{[s-1,t-1]}^n$  ($n=1,\ldots,N$).
\item[] {\bf Step 1:} 
Generate $x^n_t \sim q_{t,\theta^n}(x^n_{t} \mid x^n_{t-1},y_{t})$ and set
\begin{eqnarray}
W_{[s-1,t]}^n 
&\propto&  \frac{f_{\theta^n}(x_t^n \mid x_{t-1}^n, y_{t-1})g_{\theta^n}(y_t \mid x_t^n)}{q_{t,\theta^n}(x_t^n \mid x_{t-1}^n,y_{t})}\times
W_{[s-1,t-1]}^n.
\end{eqnarray}
\item[]{\bf Step 2:} 
Update the weight 
\begin{eqnarray}
W_{[s,t]}^n &\propto&  g_{\theta^n}(y_{s-1} \mid x_{s-1}^n,x_{s}^n)^{-1} \times W_{[s-1,t]}^n,
\end{eqnarray}
and discard $x_{s-1}^n$.
%
\end{enumerate}
\vspace{-4mm}
\hspace{0.8cm}\hrulefill
\normalsize
\end{table}

\subsection{Weight degeneracy problem}
\label{sec: simple methodology}
Using the importance weight (\ref{eq:weight-step2}) in Step 2 is obviously problematic because it would take an extremely high value when $g_{\theta}$ is close to 0. This causes the ESS to rapidly drop and triggers the MCMC update steps many times, which makes the estimation time-consuming.
Further, in Step 1, one might think it will work without any problem as long as we choose an appropriate proposal distribution $q_{t,\theta}$. However, as we shall see in illustrative examples in Section \ref{sec: Applications}, this step also causes a serious degeneracy problem. In Section \ref{sec: Marginalized block sampling tools}, we overcome this difficulty of the weight degeneracy by proposing a novel sampling method called ``a double-block sampler''with the conditional SMC update. 

\section{Particle rolling MCMC with double-block sampler}
\label{sec: Marginalized block sampling tools}
We consider sampling a block of state variables when we add the new observation or remove the old observation. For example, we update values of $\{x^n_{t-K:t-1}\}_{n=1}^N$ in addition to generating $\{x^n_t\}_{n=1}^N$ when we learn the information of $y_t$. We call this process the forward block sampling (Step 1), and the backward block sampling (Step 2) can also be defined in a similar manner. 
The double-block sampler addresses the weight degeneracy problem by reducing the path dependence between the new particle and the old particle values that are not updated.

\subsection{Idealised double-block sampler}
\label{sec:idealized double-block sampler}
We first consider the `idealised' double-block sampler where we assume an appropriate $K+1$ dimensional proposal distribution is available for the block sampling. 
In the framework of the SMC sampler, in Step 1, we have samples from the old target density $\pi(x_{s-1:t-1}, \theta\mid y_{s-1:t-1})$ and generate $x^{\dagger}_{t-K:t}$ from  the proposal kernel $\pi(x^{\dagger}_{t-K:t}|x_{s-1:t-K-1},y_{s-1:t},\theta)$. The new target density is $\pi(x_{s-1:t-K-1}, x^{\dagger}_{t-K:t}, \theta \mid y_{s-1:t})$ with the backward kernel $\pi(x_{t-K:t-1} \mid x_{s-1:t-K-1}, y_{s-1:t-1},\theta)$. In Step 2, we have samples from  the old target density $\pi(x_{s-1:t}, \theta\mid y_{s-1:t})$ and generate a candidate $x_{s:s+K-1}^{\dagger}$ using the proposal kernel   $\pi(x_{s:s+K-1}^{\dagger}|x_{s+K:t},y_{s:t},\theta)$.
Our new target density is $\pi( x_{s:s+K-1}^{\dagger}, x_{s+K:t},\theta \mid y_{s:t})$ with the backward kernel $\pi(x_{s-1:s+K-1} \mid x_{s+K:t},y_{s-1:t},\theta)$. Finally, we discard $x_{s-1}$. Details are given below.\\

\noindent
{\bf Step 1}.  We include a new observation $y_t$ in the information set, and sample from  $\pi(x_{s-1:t}, \theta \mid y_{s-1:t})$. Given the current sample $(x_{s-1:t-K-1}, \theta)$ from $\pi(x_{s-1:t-K-1}, \theta\mid y_{s-1:t-1})$, we generate $x_{t-K:t}^{\dagger} \sim \pi(x_{t-K:t}^{\dagger} \mid x_{s-1:t-K-1},y_{s-1:t},\theta)$.
Since 
\begin{eqnarray*}
\lefteqn{\pi(x_{s-1:t-K-1}, x_{t-K:t}^{\dagger},\theta|y_{s-1:t})
} && \\ 
&=&
\frac{p(x_{s-1:t-K-1}, x_{t-K:t}^{\dagger}, y_t, \theta|y_{s-1:t-1})}{p(y_t|y_{s-1:t-1})}
\\
&=&
\pi(x_{s-1:t-K-1}, \theta|y_{s-1:t-1})\times
\frac{p(x_{t-K:t}^{\dagger},y_t|x_{s-1:t-K-1},y_{s-1:t-1},\theta)}{p(y_t|y_{s-1:t-1})}
\\
&=&
\pi(x_{s-1:t-K-1}, \theta|y_{s-1:t-1})\pi(x_{t-K:t}^{\dagger}|x_{s-1:t-K-1},y_{s-1:t},\theta)\times
\frac{p(y_t|x_{s-1:t-K-1},y_{s-1:t-1},\theta)}{p(y_t|y_{s-1:t-1})}
\end{eqnarray*}
the unnormalized incremental weight is
\begin{eqnarray*}
\frac{\pi(x_{s-1:t-K-1}, x_{t-K:t}^{\dagger},\theta|y_{s-1:t})\pi(x_{t-K:t-1}|x_{s-1:t-K-1},y_{s-1:t-1},\theta)}
     {\pi(x_{s-1:t-1}, \theta|y_{s-1:t-1})\pi(x_{t-K:t}^{\dagger}|x_{s-1:t-K-1},y_{s-1:t},\theta)}
=\frac{p(y_t|x_{s-1:t-K-1},y_{s-1:t-1},\theta)}{p(y_t|y_{s-1:t-1})}.
\end{eqnarray*}
Thus we let $x_{s-1:t}=(x_{s-1:t-K-1},x_{t-K:t}^{\dagger})$ and update the importance weight as
\begin{eqnarray}
W_{[s-1,t]}^n 
&\propto&  
p(y_t|x_{t-K-1}^n,y_{s-1:t-1},\theta^n) \times W_{[s-1,t-1]}^n,
\label{incremental weight 1-1}
\end{eqnarray}
noting that $p(y_t|x_{s-1:t-K-1},y_{s-1:t-1},\theta)=p(y_t|x_{t-K-1},y_{s-1:t-1},\theta)$. \\

\noindent
{\bf Step 2}.  We remove the old observation $y_{s-1}$ from the information set, and sample from $\pi(x_{s-1:t}, \theta\mid y_{s:t})$. Given the current sample from $\pi(x_{s-1:t}, \theta|y_{s-1:t})$, we generate $x_{s:s+K-1}^{\dagger} \sim \pi(x_{s:s+K-1}^{\dagger}|x_{s+K:t},y_{s:t},\theta)$.
The unnormalized incremental weight is
\begin{eqnarray*}
\frac{\pi(x_{s:s+K-1}^{\dagger}, x_{s+K:t},\theta \mid y_{s:t})\pi(x_{s-1:s+K-1} \mid x_{s+K:t},y_{s-1:t},\theta)}
     {\pi(x_{s-1:t}, \theta|y_{s-1:t})\pi(x_{s:s+K-1}^{\dagger}|x_{s+K:t},y_{s:t},\theta)}
=
\frac{p(y_{s-1}|y_{s:t})}{p(y_{s-1}|x_{s+K:t},y_{s:t},\theta)},
\end{eqnarray*}
since 
\begin{eqnarray*}
\pi(x_{s:s+K-1}^{\dagger}, x_{s+K:t},\theta \mid y_{s:t})&=&
\pi(x_{s:s+K-1}^{\dagger}|x_{s+K:t},y_{s:t},\theta)\pi(x_{s+K:t},\theta \mid y_{s:t})
\end{eqnarray*}
and
\begin{eqnarray*}
\pi(x_{s-1:t}, \theta|y_{s-1:t}) 
&=& \pi(x_{s-1:s+K-1}|x_{s+K:t},y_{s-1:t},\theta)\pi(x_{s+K:t}, \theta|y_{s-1:t})
\\
&=&\pi(x_{s-1:s+K-1}|x_{s+K:t},y_{s-1:t},\theta) \times \frac{p(x_{s+K:t}, y_{s-1}, \theta|y_{s:t})}{p(y_{s-1}|y_{s:t})}
\\
&=& \pi(x_{s-1:s+K-1}|x_{s+K:t},y_{s-1:t},\theta)\pi(x_{s+K:t},\theta|y_{s:t})\times \frac{p(y_{s-1}|x_{s+K:t},y_{s:t},\theta)}{p(y_{s-1}|y_{s:t})}.
\end{eqnarray*}
Noting that $p(y_{s-1}\mid x_{s+K:t},y_{s:t},\theta)=p(y_{s-1} \mid x_{s+K},y_{s:t},\theta)$, we set $x_{s:t}=(x_{s:s+K-1}^{\dagger},x_{s+K:t})$ and update the importance weight
\begin{eqnarray}
W_{[s,t]}^n 
&\propto&  
p(y_{s-1}|x_{s+K}^n,y_{s:t},\theta^n)^{-1} \times W_{[s-1,t]}^n,
\label{incremental weight 2-1}
\end{eqnarray}
and discard $x_{s-1}^n$. The sampling algorithm is summarized in  Algorithm 2.\\

\begin{table}[H]
\centering
{\bf Algorithm 2: Idealised double-block sampler}
\label{table:Algorithm2_forward}\\
\footnotesize
\hspace{0.4cm}
\hrulefill
\vspace{-2mm}
\\
\begin{enumerate}
\item[]
Let  $(x_{s-1:t-1}^n,\theta^n)$ denote the sample from $\pi(x_{s-1:t-1}, \theta \mid y_{s-1:t-1})$ with the weight $W_{[s-1,t-1]}^n$  ($n=1,\ldots,N$).
\item[] {\bf Step 1:} 
Generate $x_{t-K:t}^n \sim \pi(x_{t-K:t}^n \mid x_{s-1:t-K-1}^n,y_{s-1:t},\theta^n)$ and set
\begin{eqnarray}
W_{[s-1,t]}^n 
&\propto&  
p(y_t|x_{t-K-1}^n,y_{s-1:t-1},\theta^n) \times W_{[s-1,t-1]}^n.
\end{eqnarray}
\item[] {\bf Step 2}. 
Generate $x_{s:s+K-1}^n \sim \pi(x_{s-1:s+K-1}^n \mid x_{s+K:t}^n,y_{s:t},\theta^n)$ and set
\begin{eqnarray}
W_{[s,t]}^n  & \propto &  p(y_{s-1} \mid x_{s+K}^n,y_{s:t},\theta^n)^{-1} \times W_{[s-1,t]}^n,
\end{eqnarray}
and discard $x_{s-1}^n$.
%
%
\vspace{-4mm}
\end{enumerate}
\hspace{0.8cm}\hrulefill
\normalsize
\end{table}

\subsection{Practical double-block sampler}
In practice, it is often difficult to find an `idealised' proposal distribution for the block sampling. Hence, we adopt the approach of the conditional SMC update for the particle Gibbs sampler  (\shortciteN{AndrieuDoucetHolenstein(10)}), which considers the artificial target density ad generates a cloud of values for one particle path. 
%
%
%
%
%
%

In Step 1, we have samples from  the old target density $\pi(x_{s-1:t-1}, \theta \mid y_{s-1:t-1})$, and generate the indices $k_{t-K:t-1}=(k_{t-K},\ldots,k_{t-1})$ and a cloud of particles from the proposal kernel  $\psi_\theta$ defined in (\ref{Psi in the filtering}). The new target density is $\pi(x_{s-1:t-K-1}, x^{\dagger}_{t-K:t}, \theta \mid y_{s-1:t})$ with the backward kernel $\hat{\pi}/\pi(x_{s-1:t-K-1}, x^{\dagger}_{t-K:t}, \theta \mid y_{s-1:t})$ where $\hat{\pi}$ is defined in (\ref{target dist in Block sampling in augmented space})\footnote{The marginal density of $\hat{\pi}$ is $\pi(x_{s-1:t-K-1}, x^{\dagger}_{t-K:t}, \theta \mid y_{s-1:t})$ as shown in Proposition \ref{prop 2} with $x^{\dagger}_{t-K:t}=(x_{t-K}^{k_{t-K}^*},\ldots,x_t^{k_t^*})$.}. We set $x_{s-1:t}=\left(x_{s-1:t-K-1}, x^{\dagger}_{t-K:t}\right)$ which is the sample from the new target density with the unnormalized incremental weight  $\hat{p}(y_t \mid x_{t-K-1}^n,y_{s-1:t-1},\theta^n)$ in (\ref{eq:phat_forward}).

In Step 2, we have samples from  the old target density $\pi(x_{s-1:t}, \theta\mid y_{s-1:t})$, and generate the indices $k_{s-1:s+K-1}$ and a cloud of particles from the proposal kernel  $\bar{\psi}_\theta$ defined in (\ref{Psi in the forgetting}). The new target density is $\pi(x_{s:s+K-1}^{\dagger}, x_{s+K:t}, \theta \mid y_{s:t})$ with the backward kernel $\check{\pi}/\pi(x_{s:s+K-1}^{\dagger}, x_{s+K:t}, \theta \mid y_{s:t})$ where $\check{\pi}$ defined in (\ref{target dist in rolling est})\footnote{The marginal density of $\check{\pi}$ is $\pi(x_{s:s+K-1}^{\dagger}, x_{s+K:t}, \theta \mid y_{s:t})$ as shown in Proposition \ref{prop in discarding steps} with $x^{\dagger}_{s:s+K-1}=(x_{s}^{k_{s}^*},\ldots,x_{s+K-1}^{k_{s+K-1}^*})$.}. We set $x_{s:t}=\left(x_{s:s+K-1}^{\dagger}, x_{s+K:t}\right)$ which is the sample from the new target density with the unnormalized incremental weight  $\hat{p}(y_{s-1} \mid x_{s+K}^n,y_{s:t},\theta^n)^{-1}$ in (\ref{def of approx of conditinal likelihood 2}).
Details are given below.
\subsubsection{Forward block sampling (Step 1)}
\label{sec: MBS to incorporate the new information}

We first generate a number of candidates $x_{t-K:t-1}^{n,m}$ $(m=1,\ldots,M)$ with the current values $x_{t-K:t-1}^n$ fixed using the conditional SMC. Then, for each $x_{t-K:t-1}^{n,m}$, we generate $x^{n,m}_t$.
In this `local particle filtering', we resample the particles at $t-K+1,\dots,t$. This operation is equivalent to choosing the `parent' $x_{j}^{n,m}$ for $x_{j+1}^{n,m}$ $(j=t-K,\dots,t-1$). Using this terminology, if we choose one particle $x_t^{n,m}$, its `ancestors' are uniquely determined from $x_{j}^{n,m}$ ($j = t-K,\dots,t-1$). We call this descendant and its ancestors the `lineage'. In the conditional SMC step, fixing the current values $x_{t-K:t-1}^n$ is seen as fixing one lineage by choosing their indices $k_j\ (j=t-K,\dots,t-1)$  (where we drop the superscript $n$ for simplicity) which follows the rule
\begin{align}
a^{k_{j+1}}_{j} &= k_{j},\quad j = t-K,\dots,t-1. \label{rule on indices}
\end{align}
In addition, the index of their descendant is determined as $k_t = 1$.

After generating $x_{t-K:t}^{n,m}$ $(m=1,\ldots,M)$, we choose one lineage to store as the next values of $x^n_{t-K:t}$. This is equivalent to sampling a random index $k^*_t$ for the candidate $x^{n,m}_t$ and identifying the ancestors for which indices are obtained by following the rule
\begin{eqnarray}
a^{k^*_{j+1}}_{j} &=& k^*_{j}, \quad j = t-K,\dots,t-1 \label{rule on indices star}.
\end{eqnarray}
Moreover, we can improve its efficiency by implementing `smoothing' for the generated candidates following the algorithm reported in \shortciteN{WhiteleyAndrieuDoucet(10)}.
In this smoothing step, we again choose $k^*_j$ for $j=t-K,\dots,t-1$ randomly. 
This manipulation of breaking the relationship between the parent and the child in the lineage is effective in improving the mixing property, or sampling values of $x^n_{t-K:t-1}$ that may be different from the lineages obtained in the previous step.

\begin{figure}[H]
\centering
\caption{Forward block sampling and particle simulation smoother.}
\label{figure:forward_diagram}
\includegraphics[width=9cm]{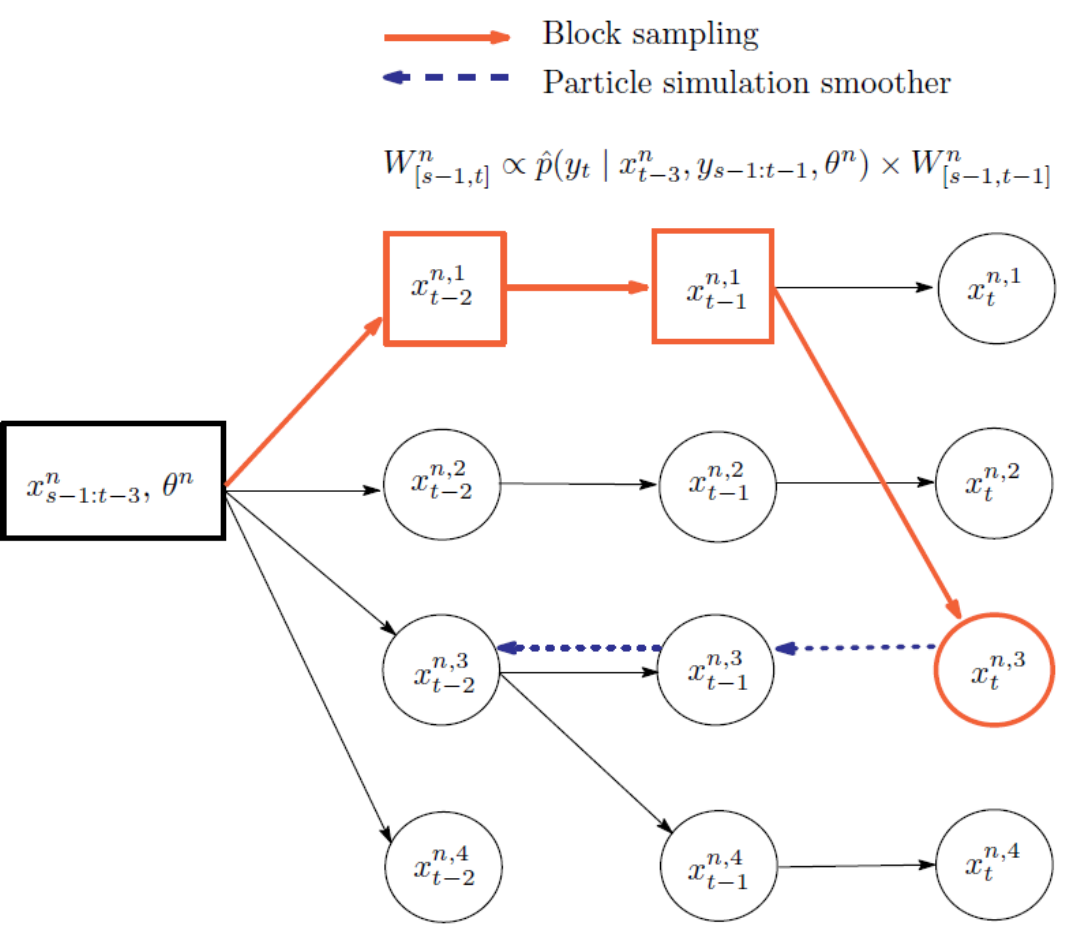}
\end{figure}

The detailed algorithm is provided below. We fix one lineage in Step 1-1(a) and implement the conditional SMC in Steps 1-1(b) and 1-1(c). The candidates for $x^n_t$ are generated in Step 1-1(d) and we compute the importance weight for the $n$-th particle in the `global particle filtering' in Step 1-2. The smoothing is implemented in Step 1-3. 
\begin{enumerate}
\item We generate $x_{t-K:t}^n \sim \pi(x_{t-K:t}^n \mid x_{s-1:t-K-1}^n,y_{s-1:t},\theta^n)$ using the conditional SMC update:
\begin{enumerate}
\item[(a)] Sample $k_j$ from $\{1,\ldots,M\}$ with probability $1/M$ ($j = t-K,\dots,t-1$)  and set 
\begin{eqnarray*}
(x^{n,k_{t-K}}_{t-K},\dots,x^{n,k_{t-1}}_{t-1}) =  x_{t-K:t-1}^n,\quad
(a^{k_{t-K+1}}_{t-K},\dots,a^{k_{t}}_{t-1})  =  (k_{t-K},\dots,k_{t-1}), 
\end{eqnarray*}
where $x_{t-K:t-1}^n$ is a current sample with the importance weight $W_{[s-1,t-1]}^n$.

\item [(b)]
Set $x^{n,a^m_{t-K-1}}_{t-K-1} = x_{t-K-1}^n$ for all $m$ according to the convention, and sample $x^{n,m}_{t-K} \sim q_{t-K,\theta^n}(\cdot \mid x_{t-K-1}^{n},y_{t-K})$ for each $m\in \{1,\dots,M\}\setminus \{k_{t-K}\}$. Let $j=t-K+1$. 

\item[(c)]

Sample $a^{m}_{j-1} \sim \mathcal{M}(V^{1:M}_{j-1,\theta^n})$
and $x^{n,m}_j \sim q_{j,\theta^n}(\cdot \mid x_{j-1}^{n,a^m_{j-1}},y_{j})$ for each $m\in \{1,\dots,M\}\setminus \{k_j\}$
where $V^{1:M}_{j-1,\theta^n} \equiv (V^{1}_{j-1,\theta^n},\ldots,V^{M}_{j-1,\theta^n})$ and
\begin{eqnarray}
V^{m}_{j,\theta^n} &=& \frac{v_{j,{\theta}^n}(x^{n,a^m_{j-1}}_{j-1},x_{j}^{n,m})}{\displaystyle \sum^M_{i=1}v_{j,{\theta}^n}(x^{n,a^i_{j-1}}_{j-1},x_j^{n,i})}, 
\label{eq:V}
\\
&&
\hspace{-6mm}
v_{j,\theta^n}(x^{n,a^{m}_{j-1}}_{j-1},x_{j}^{n,m}) = \frac{f_{\theta^n}(x^{n,m}_{j} \mid x^{n,a^m_{j-1}}_{j-1},y_{j-1})g_{\theta^n}(y_{j} \mid x_{j}^{n,m})}{q_{j,{\theta}^n}(x_{j}^{n,m} \mid x_{j-1}^{n,a^m_{j-1}},y_{j})}, \quad m=1,\ldots,M.\nonumber\\
&&
\end{eqnarray}

\item[(d)] 
If $j <t-1$, set $j\leftarrow j+1$ and go to (c). Otherwise,  
sample $x^{n,m}_t$ $(m=1,\ldots,M)$ and $k_t^*$ as follows.
\begin{enumerate}
\item[(i)] Sample $x^{n,1}_t \sim q_{t,\theta^n}(\cdot \mid x^{n,k_{t-1}}_{t-1},y_t)$.
\item[(ii)] Sample $a^{n,m}_{t-1} \sim \mathcal{M}(V^{1:M}_{t-1,\theta^n})$ and $x^{n,m}_t \sim q_{t,\theta^n}(\cdot \mid x_{t-1}^{n,a^m_{t-1}},y_{t})$ for each $m\in \{2,\dots,M\}$.
\item[(iii)] Sample $k^*_t \sim \mathcal{M}(V^{1:M}_{t,\theta^n})$ and obtain  $k^*_j\ (j=t-1,\dots,t-K)$ using  (\ref{rule on indices star}). 
\end{enumerate}
\end{enumerate}
\item Let $x_{s-1:t}^n=(x_{s-1}^{n},\ldots,x_{t-K-1}^{n}, x_{t-K}^{n,k_{t-K}^*},\ldots,x_{t}^{n,k_t^*})$ and compute the importance weight\footnote{we use the notation $\hat{p}(y_t \mid x_{t-K-1}^n,y_{s-1:t-1},\theta^n)$ since it is an unbiased estimator of $p(y_t \mid x_{t-K-1}^n,y_{s-1:t-1},\theta^n)$ as we shall show in Proposition \ref{decomp of variance of incremental weight}.}
\begin{eqnarray}
W^n_{[s-1,t]} &\propto& \hat{p}(y_t \mid x_{t-K-1}^n,y_{s-1:t-1},\theta^n)\times W^n_{[s-1,t-1]}. 
\label{incremental weight in the plain MBS}
\\
&& \hat{p}(y_t \mid x_{t-K-1}^n,y_{s-1:t-1},\theta^n)=\frac{1}{M}\sum^M_{m=1}v_{t,\theta^n}(x^{n,a^{m}_{t-1}}_{t-1},x_t^{n,m}),
\label{eq:phat_forward}
\end{eqnarray}
where $\hat{p}(y_t \mid x_{t-K-1}^n,y_{s-1:t-1},\theta^n)$ can be seen as the estimate of the intractable incremental weight $p(y_t \mid x_{t-K-1}^n,y_{s-1:t-1},\theta^n)$  in (\ref{incremental weight 1-1}) for the idealised double-block sampler. 

\item Implement the particle simulation smoother to sample $(k^*_{t-K},k^*_{t-K+1},\dots,k^*_t)$ jointly. Generate $k^*_{j}\sim \mathcal{M}(\bar{V}_{j,\theta}^{1:M}),$ $j=t-1,\dots,t-K$, recursively where
\begin{eqnarray}
\label{eq:vbar}
\bar{V}^{m}_{j,\theta}
\equiv
\frac{V^{m}_{j,\theta}f_\theta(x^{k^*_{j+1}}_{j+1} \mid x^{m}_{j},y_{j+1})}
{\sum_{i=1}^MV^{i}_{j,\theta}f_\theta(x^{k^*_{j+1}}_{j+1} \mid x^{i}_{j},y_{j+1})}, \quad m=1,\ldots,M,
\end{eqnarray}
and set $x_{s-1:t}^n=(x_{s-1}^{n},\ldots,x_{t-K-1}^{n}, x_{t-K}^{n,k_{t-K}^*},\ldots,x_{t}^{n,k_t^*})$.
\end{enumerate}
\begin{table}[H]
\centering
{\bf Algorithm 3 (Step 1) : Practical double-block sampler}
\label{table:Algorithm3_forward}\\
\footnotesize
\hspace{0.4cm}
\hrulefill
\vspace{-2mm}
\\
\begin{enumerate}
\item[]
Let  $(x_{s-1:t-1}^n,\theta^n)$ denote the sample from $\pi(x_{s-1:t-1}, \theta \mid y_{s-1:t-1})$ with the weight $W_{[s-1,t-1]}^n$  ($n=1,\ldots,N$).
\item[] 1.
We generate $x_{t-K:t}^n \sim \pi(x_{t-K:t}^n \mid x_{s-1:t-K-1}^n,y_{s-1:t},\theta^n)$ using the conditional SMC update:
\begin{enumerate}
\item[(a)]
Sample $k_j$ from $\{1,\ldots,M\}$ with probability $1/M$ ($j = t-K,\dots,t-1$)  and set 
\begin{eqnarray*}
(x^{n,k_{t-K}}_{t-K},\dots,x^{n,k_{t-1}}_{t-1}) =  x_{t-K:t-1}^n,\quad
(a^{k_{t-K+1}}_{t-K},\dots,a^{k_{t}}_{t-1})  =  (k_{t-K},\dots,k_{t-1}). 
\end{eqnarray*}
\item[(b)]
Set $x^{n,a^m_{t-K-1}}_{t-K-1} = x_{t-K-1}^n$ and sample $x^{n,m}_{t-K} \sim q_{t-K,\theta^n}(\cdot \mid x_{t-K-1}^{n},y_{t-K})$, $m\in \{1,\dots,M\}\setminus \{k_{t-K}\}$. Let $j=t-K+1$. 
\item[(c)]
Sample $a^{m}_{j-1} \sim \mathcal{M}(V^{1:M}_{j-1,\theta^n})$
and $x^{n,m}_j \sim q_{j,\theta^n}(\cdot \mid x_{j-1}^{n,a^m_{j-1}},y_{j})$, $m\in \{1,\dots,M\}\setminus \{k_j\}$
where $V^{1:M}_{j-1,\theta^n}$ is given by (\ref{eq:V}).
\item[(d)] 
If $j <t-1$, set $j\leftarrow j+1$ and go to (c). Otherwise,  
sample $x^{n,m}_t$ $(m=1,\ldots,M)$ and $k_t^*$ as follows.
\begin{enumerate}
\item[(i)] Sample $x^{n,1}_t \sim q_{t,\theta^n}(\cdot \mid x^{n,k_{t-1}}_{t-1},y_t)$.
\item[(ii)] Sample $a^{n,m}_{t-1} \sim \mathcal{M}(V^{1:M}_{t-1,\theta^n})$ and $x^{n,m}_t \sim q_{t,\theta^n}(\cdot \mid x_{t-1}^{n,a^m_{t-1}},y_{t})$ for $m\in \{2,\dots,M\}$.
\item[(iii)] Sample $k^*_t \sim \mathcal{M}(V^{1:M}_{t,\theta^n})$ and obtain  $k^*_j\ (j=t-1,\dots,t-K)$ using  (\ref{rule on indices star}). 
\end{enumerate}
\end{enumerate}
\item[] 2. 
Let $x_{s-1:t}^n=(x_{s-1}^{n},\ldots,x_{t-K-1}^{n}, x_{t-K}^{n,k_{t-K}^*},\ldots,x_{t}^{n,k_t^*})$ and update the weight
\begin{eqnarray*}
W^n_{[s-1,t]} &\propto& \hat{p}(y_t \mid x_{t-K-1}^n,y_{s-1:t-1},\theta^n)\times W^n_{[s-1,t-1]}. 
\end{eqnarray*}
where $\hat{p}$ is defined in (\ref{eq:phat_forward}).
\item[] 3. The particle simulation smoother. Generate $k^*_{j}\sim \mathcal{M}(\bar{V}_{j,\theta}^{1:M}),$ $j=t-1,\dots,t-K$, where $\bar{V}^{m}_{j,\theta}$ is given in (\ref{eq:vbar}), and set $x_{s-1:t}^n=(x_{s-1}^{n},\ldots,x_{t-K-1}^{n}, x_{t-K}^{n,k_{t-K}^*},\ldots,x_{t}^{n,k_t^*})$.
\end{enumerate}
\vspace{-4mm}
\hspace{0.8cm}\hrulefill
\normalsize
\end{table}
Figure \ref{figure:forward_diagram} illustrates an example with $K=2$, $M=4$ and the current sample $(x_{s-1:t-1}^n,\theta^n)$.
\begin{enumerate}
\item
\begin{enumerate}
\item[(a)] Sample $k_{t-2}$ and $k_{t-1}$ from $\{1,2,3,4\}$ with probability 1/4 and suppose $k_{t-2}=k_{t-1}=1$. 
We set $x_{t-2}^{n,1}=x_{t-2}^n$, $x_{t-1}^{n,1}=x_{t-1}^n$ (with the red rectangle) and
$(a_{t-2}^{1},a_{t-1}^{1})=(1,1)$.
\item[(b)] Set $x^{n,a^m_{t-3}}_{t-3} = x_{t-3}^n$ for all $m$ (with the black rectangle), and sample $x^{n,m}_{t-2} \sim q_{t-2,\theta^n}(\cdot \mid x_{t-3}^{n},y_{t-2})$ for each $m\in \{2,3,4\}$ (with the black circle). 
\item[(c)]
Sample $a^{m}_{t-2} \sim \mathcal{M}(V^{1:4}_{t-2,\theta^n})$  for $m\in \{2,3,4\}$ and suppose $a^{2}_{t-2}=2$, $a^{3}_{t-2}=3$, $a^{4}_{t-2}=3$. Generate 
 $x^{n,m}_{t-1} \sim q_{t-1,\theta^n}(\cdot \mid x_{t-2}^{n,a^m_{t-2}},y_{t-1})$ for  $m\in \{2,3,4\}$ (with the black circle).

\item[(d)] 
\begin{enumerate}
\item[(i)] Sample $x^{n,1}_t \sim q_{t,\theta^n}(\cdot \mid x^{n,1}_{t-1},y_t)$.
\item[(ii)] Sample $a^{n,m}_{t-1} \sim \mathcal{M}(V^{1:4}_{t-1,\theta^n})$  for $m\in \{2,3,4\}$ and suppose $a^{2}_{t-1}=2$, $a^{3}_{t-1}=1$, $a^{4}_{t-1}=4$. Generate and $x^{n,m}_t \sim q_{t,\theta^n}(\cdot \mid x_{t-1}^{n,a^m_{t-1}},y_{t})$ for $m\in \{2,3,4\}$.
\item[(iii)] Sample $k^*_t \sim \mathcal{M}(V^{1:4}_{t,\theta^n})$ and suppose $k^*_t=3$. Using (\ref{rule on indices star}), we obtain $k_{t-1}^*=k_{t-2}^*=1$ and select ($x_{t}^{n,3},x_{t-1}^{n,1},x_{t-2}^{n,1})$ with red lines.
\end{enumerate}
\end{enumerate}
\item  Let $x_{s-1:t}^n=(x_{s-1}^{n},\ldots,x_{t-3}^{n}, x_{t-2}^{n,1},x_{t-1}^{n,1},x_{t}^{n,3})$ and compute the importance weight.
\item Implement the particle simulation smoother to sample $(k^*_{t-2},k^*_{t-1},k^*_t)$ jointly. Generate $k^*_{j}\sim \mathcal{M}(\bar{V}_{j,\theta}^{1:4}),$ $j=t-1,t-2$, recursively (with dotted lines) and suppose $k^*_{t-1}=3$ and $k^*_{t-2}=3$. We set $x_{s-1:t}^n=(x_{s-1}^{n},\ldots,x_{t-3}^{n}, x_{t-2}^{n,3}, x_{t-1}^{n,3},x_{t}^{n,3})$.
\end{enumerate}

\noindent
\remark
As the proposal density $q_{j,\theta}$, we can either use the prior density $f_\theta$ or more sophisticated density that incorporates the information of the likelihood $g_\theta$. Even if we use the prior $f_\theta$ as the proposal, the above sampling becomes much more efficient than the simple rolling-window sampler as shown in Section \ref{sec: Applications}. 
\subsubsection{Backward block sampling (Step 2)}
\label{sec: Algorithm: inverted marginal block sampling scheme}
Before we describe the backward block sampling which generates a cloud of particles based on $(x^n_{s+K:t},\theta^n)$, we define the notation for the particle index as noted in the forward block sampling but in the reverse order. A `parent' particle of $x^m_{j}$ is chosen from $x^{1:M}_{j+1}$ (not from $x^{1:M}_{j-1}$) and consequently $a^m_{j+1}$ denotes its parent's index. In this case, the relationship of $a^m_{j+1}$ and $k_{j}$ is given as follows:
\begin{eqnarray}
\label{eq:index_backward}
a^{k_{j}}_{j+1} = k_{j+1}, \quad j=s+K-2,\dots,s-2.
\end{eqnarray}
For each $n$, we first generate $M$ particle paths, $x_{s-1:s+K-1}^{n,1:M}\equiv(x_{s-1:s+K-1}^{n,1},\ldots,x_{s-1:s+K-1}^{n,M})$, and sample one path, $x_{s:t}^n$,  from $x_{s:s+K-1}^{n,1:M}$ as noted below.  

\begin{enumerate}
\item We generate $x_{s-1:s+K-1}^n \sim \pi(x_{s-1:s+K-1}^n \mid x_{s+K:t}^n,y_{s:t},\theta^n)$.
\begin{enumerate}
\item[(a)] Sample indices $k_j$ from $\{1,\ldots,M\}$ with probability $1/M$ ($j = s+K-1,s+K-2,\dots,s-1$) and set
\begin{eqnarray*}
(x^{n,k_{s-1}}_{s-1},\dots,x^{n,k_{s+K-1}}_{s+K-1}) =  x_{s-1:s+K-1}^n,\quad
(a^{k_{s-2}}_{s-1},\dots,a^{k_{s+K-2}}_{s+K-1})  =  (k_{s-1},\dots,k_{s+K-1}), 
\end{eqnarray*}
where $x_{s-1:s+K-1}^n$ is a current sample with the importance weight $W_{[s-1,t]}^n$.

\item[(b)] Set $x_{s+K}^{n,a_{s+K}^m}=x_{s+K}^n$ for all $m$ according to the convention, and sample $x^{n,m}_{s+K-1} \sim q_{s+K-1,\theta^n}(\cdot \mid x_{s+K}^n, y_{s+K-1})$
for each $m\in \{1,\dots,M\}\setminus \{k_{s+K-1}\}$. Let $j=s+K-2$.
\item[(c)]
Sample $a^{m}_{j+1} \sim \mathcal{M}(V^{1:M}_{j+1,\theta^n})$ and $x^{n,m}_{j} \sim q_{j,\theta^n}(\cdot \mid x_{j+1}^{n,a^m_{j+1}}, y_{j})$ for each $m\in \{1,\dots,M\}\setminus \{k_{j}\}$ where
$V^{1:M}_{j+1,\theta^n}=(V^{1}_{j+1,\theta^n},\ldots, V^{M}_{j+1,\theta^n})$ and
\begin{align}
\label{eq:back_V}
&V^{m}_{j,\theta^n}=\frac{v_{j,\theta^n}(x_{j}^{n,m},x^{n,a^{m}_{j+1}}_{j+1})}{\sum^M_{i=1}v_{j,\theta^n}(x_{j}^{n,i},x^{n,a^{i}_{j+1}}_{j+1})},  \\
\label{eq:back_v}
&v_{j,\theta^n}(x_{j}^{n,m},x^{n,a^{m}_{j+1}}_{j+1}) = \frac{p(x^{n,m}_{j} \mid x^{n,a^m_{j+1}}_{j+1},\theta)g_{\theta^n}(y_{j} \mid x_{j}^{n,m},x^{n,a^m_{j+1}}_{j+1})}{q_{j,{\theta}^n}(x_{j}^{n,m} \mid x_{j+1}^{n,a^m_{j+1}},y_{j})}, \quad m=1,\ldots,M.
\end{align}
\item[(d)] If $j>s-1$, set $j\leftarrow j-1$ and go to (c)\footnote{Note that we need to generate $x_{s-1}^{n,m}$ to compute $\hat{p}$ in (\ref{def of approx of conditinal likelihood 2}).}. Otherwise, 
sample $k^{*}_{s} \sim \mathcal{M}(V^{1:M}_{s,\theta^n})$ and obtain $k_j^*$ $(j=s+1,\ldots,s+K-1)$ using (\ref{eq:index_backward}).
%
%
\end{enumerate}
\item Let $x_{s:t}^n=(x_s^{n,k_s^*}, \ldots,x_{s+K-1}^{n,k_{s+K-1}^*},x_{s+K}^n,\ldots,x_t^n)$  and compute its importance weight
\begin{eqnarray}
\label{eq:weight_backward}
W^n_{[s,t]} \propto 
\left\{
\begin{array}{ll}
\frac{1}{\hat{p}(y_{s-1} \mid x^n_{s+K},y_{s:t},\theta^n)}W^n_{[s-1,t]}, & \mbox{if $\hat{p}(y_{s-1} \mid x_{s+K}^n,y_{s:t},\theta^n)\neq 0$}, \\
0, & \mbox{if $\hat{p}(y_{s-1} \mid x_{s+K}^n,y_{s:t},\theta^n)=0$},
\end{array}
\right.
\end{eqnarray}
where
\begin{eqnarray}
\hat{p}(y_{s-1} \mid x_{s+K}^n,y_{s:t},\theta^n) &=& \frac{1}{M}\sum^M_{m=1}v_{s-1,\theta^n}(x_{s-1}^{n,m},x_{s}^{n,a^m_{s}}),
\label{def of approx of conditinal likelihood 2}
\end{eqnarray}
and $\hat{p}(y_{s-1} \mid x_{s+K}^n,y_{s:t},\theta^n)^{-1}$ can be seen as the estimate of the intractable incremental weight $p(y_{s-1} \mid x_{s+K}^n,y_{s:t},\theta^n)^{-1}$  in (\ref{incremental weight 2-1}) for the idealised double-block sampler.

\item Implement the particle simulation smoother to sample $(k^*_{s},k^*_{s+1},\dots,k^*_{s+K-1})$ jointly. Generate $k^*_{j}\sim \mathcal{M}(\bar{V}_{j,\theta^n}^{1:M}),$ $j=s+1,\dots,s+K-1$, recursively where
\begin{eqnarray}
\label{eq:vbar_back}
\bar{V}_{j,\theta^n}^m 
=
\frac{V_{j,\theta^n}^{m}p(x_{j-1}^{k_{j-1}^*} \mid x_{j}^m,\theta^n)}
     {\sum_{i=1}^MV_{j,\theta^n}^{i}p(x_{j-1}^{k_{j-1}^*} \mid x_{j}^i,\theta^n)},
 \quad m=1,\ldots,M.
\end{eqnarray}
and set $x_{s:t}^n=(x_s^{n,k_s^*}, \ldots,x_{s+K-1}^{n,k_{s+K-1}^*},x_{s+K}^n,\ldots,x_t^n)$.
\end{enumerate}

\begin{table}[H]
\centering
{\bf Algorithm 3 (Step 2) : Practical double-block sampler}
\label{table:Algorithm3_backward}\\
\footnotesize
\hspace{0.4cm}
\hrulefill
\vspace{-2mm}
\\
\begin{enumerate}
\item[]
Let  $(x_{s-1:t}^n,\theta^n)$ denote the sample from $\pi(x_{s-1:t}, \theta \mid y_{s-1:t})$ with the weight $W_{[s-1,t]}^n$  ($n=1,\ldots,N$).
\item[] 1.
We generate $x_{s-1:s+K-1}^n \sim \pi(x_{s-1:s+K-1}^n \mid x_{s+K:t}^n,y_{s:t},\theta^n)$ as follows:
\begin{enumerate}
\item[(a)]
Sample $k_j$ from $\{1,\ldots,M\}$ with probability $1/M$ ($j = s+K-1,\dots,s-1$) and set
\begin{eqnarray*}
(x^{n,k_{s-1}}_{s-1},\dots,x^{n,k_{s+K-1}}_{s+K-1}) =  x_{s-1:s+K-1}^n,\quad
(a^{k_{s-2}}_{s-1},\dots,a^{k_{s+K-2}}_{s+K-1})  =  (k_{s-1},\dots,k_{s+K-1}). 
\end{eqnarray*}
\item[(b)]
Set $x_{s+K}^{n,a_{s+K}^m}=x_{s+K}^n$ and sample $x^{n,m}_{s+K-1} \sim q_{s+K-1,\theta^n}(\cdot \mid x_{s+K}^n, y_{s+K-1})$, $m\in \{1,\dots,M\}\setminus \{k_{s+K-1}\}$. Let $j=s+K-2$.
\item[(c)]
Sample $a^{m}_{j+1} \sim \mathcal{M}(V^{1:M}_{j+1,\theta^n})$ and $x^{n,m}_{j} \sim q_{j,\theta^n}(\cdot \mid x_{j+1}^{n,a^m_{j+1}}, y_{j})$, $m\in \{1,\dots,M\}\setminus \{k_{j}\}$ where
$V^{1:M}_{j+1,\theta^n}$ is given by (\ref{eq:back_V}).
\item[(d)] 
If $j>s-1$, set $j\leftarrow j-1$ and go to (c). Otherwise, 
sample $k^{*}_{s} \sim \mathcal{M}(V^{1:M}_{s,\theta^n})$ and obtain $k_j^*$ $(j=s+1,\ldots,s+K-1)$ using (\ref{eq:index_backward}).
\end{enumerate}
\item[] 2. 
Let $x_{s:t}^n=(x_s^{n,k_s^*}, \ldots,x_{s+K-1}^{n,k_{s+K-1}^*},x_{s+K}^n,\ldots,x_t^n)$  and update the weight
\begin{eqnarray}
W^n_{[s,t]} \propto 
\hat{p}(y_{s-1} \mid x^n_{s+K},y_{s:t},\theta^n)^{-1}\times W^n_{[s-1,t]},
\end{eqnarray}
where $\hat{p}$ is defined in (\ref{def of approx of conditinal likelihood 2}). If $\hat{p}=0$, set $W^n_{[s,t]}=0$. 
\item[] 3. The particle simulation smoother.
Generate $k^*_{j}\sim \mathcal{M}(\bar{V}_{j,\theta^n}^{1:M}),$ $j=s+1,\dots,s+K-1$, where
$\bar{V}_{j,\theta^n}^m$ is given in  (\ref{eq:vbar_back}), and set $x_{s:t}^n=(x_s^{n,k_s^*}, \ldots,x_{s+K-1}^{n,k_{s+K-1}^*},x_{s+K}^n,\ldots,x_t^n)$.
\end{enumerate}
\vspace{-4mm}
\hspace{0.4cm}\hrulefill
\normalsize
\end{table}
\begin{figure}[H]
\centering
\caption{Backward block sampling and particle simulation smoother.}
\label{figure:backward_diagram}
\includegraphics[width=9cm]{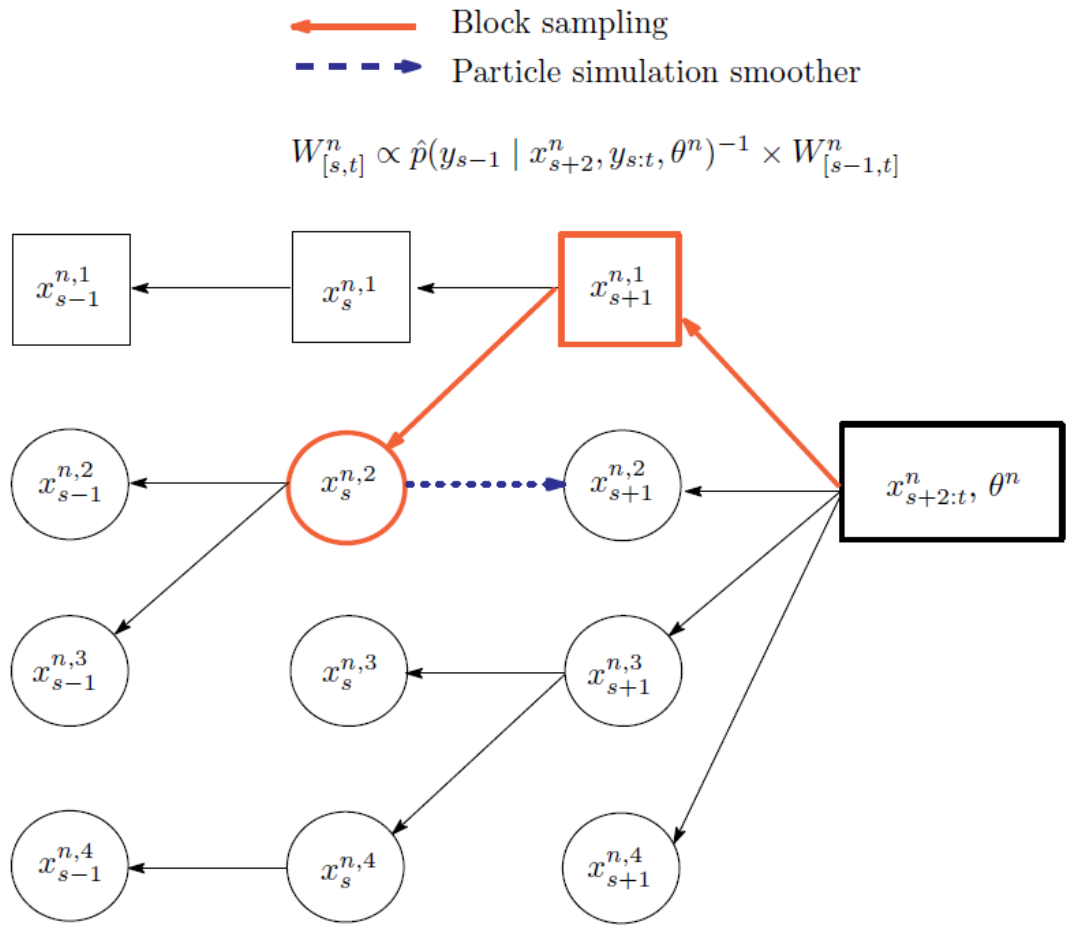}
\end{figure}

Figure \ref{figure:backward_diagram} illustrates an example with $K=2$, $M=4$ and the current sample $(x_{s-1:t}^n,\theta^n)$.

\begin{enumerate}
\item
\begin{enumerate}
\item[(a)] Sample indices $k_{s+1}$, $k_s$, $k_{s-1}$ from $\{1,2,3,4\}$ with probability $1/4$ and suppose 
$k_{s+1}=1$, $k_s=1$, $k_{s-1}=1$. We set
$(x^{n,1}_{s-1}, x^{n,1}_{s},x^{n,1}_{s+1}) =  x_{s-1:s+1}^n$ (with the rectangle) 
and $(a^{1}_{s-1},a^{1}_{s},a^{1}_{s+1})  =  (1,1,1)$. 
\item[(b)] Set $x_{s+2}^{n,a_{s+2}^m}=x_{s+2}^n$ for all $m$ (with the thick black rectangle), and sample $x^{n,m}_{s+1} \sim q_{s+1,\theta^n}(\cdot \mid x_{s+1}^n, y_{s+1})$
for $m\in \{2,3,4\}$ (with the black circle).
\item[(c)]
Sample $a^{m}_{s+1} \sim \mathcal{M}(V^{1:4}_{s+1,\theta^n})$ and suppose $a^2_{s+1}=1$, $a^3_{s+1}=3$, $a^4_{s+1}=3$.
Generate $x^{n,m}_{s} \sim q_{s,\theta^n}(\cdot \mid x_{s+1}^{n,a^m_{s+1}}, y_{s})$ for $m\in \{2,3,4\}$.

\item[(d)]
Sample $a^{m}_{s} \sim \mathcal{M}(V^{1:4}_{s,\theta^n})$ and suppose $a^2_{s}=2$, $a^3_{s}=2$, $a^4_{s}=4$.
Generate $x^{n,m}_{s-1} \sim q_{s-1,\theta^n}(\cdot \mid x_{s}^{n,a^m_{s}}, y_{s-1})$ for $m\in \{2,3,4\}$.
\item[(e)] Sample $k^{*}_{s} \sim \mathcal{M}(V^{1:4}_{s,\theta^n})$ and suppose $k^*_s=2$. Using (\ref{eq:index_backward}), we obtain $k_{s+1}^*=1$, and select ($x_{s}^{n,2},x_{s+1}^{n,1})$ with red lines.
\end{enumerate}
\item Let $x_{s:t}^n=(x_s^{n,2}, x_{s+1}^{n,1},x_{s+2}^n,\ldots,x_t^n)$  and compute its importance weight.
\item Implement the particle simulation smoother to sample $(k^*_{s},k^*_{s+1})$ jointly. Generate $k^*_{s+1}\sim \mathcal{M}(\bar{V}_{s+1,\theta^n}^{1:4}),$ and suppose $k^*_{s+1}=2$. We set $x_{s:t}^n=(x_s^{n,2}, x_{s+1}^{n,2},x_{s+2}^n,\ldots,x_t^n)$.
\end{enumerate}
\noindent \remark In Algorithm 3, we assume we can evaluate $p(x_{j-1} \mid x_j,\theta)$ given in (\ref{eq:initial_density}).\\

\noindent \remark In the simple rolling-window sampler, we reweighted the particles according to the likelihood $g_{\theta}(y_{s-1}|x_{s-1},x_s)$ in Step 2, while the unbiased estimate of the conditional likelihood $\hat{p}(y_{s-1} \mid x^n_{s+K},y_{s:t},\theta^n)$ is used in the practical double-block sampler. Algorithm 3 substantially improves the weight degeneracy since we condition on $y_{s:t}$ and integrate out $(x_{s-1},\ldots,x_{s+K-1})$.
\subsection{Sequential MCMC estimation without rolling the window}
\label{sec: Preparation for rolling window estimation}
In the above discussion, it is implicitly assumed that the initial particles  approximating $\pi(x_{1:L+1},\theta \mid y_{1:L+1})$ are obtained.
To sample from this initial posterior distribution, using MCMC-based methods is straightforward as in the warm-up period for the practical filtering described in \shortciteN{Polson2008}. 
Moreover, we could simply  use MCMC samples from the initial posterior distribution.
However, based on our proposed method for the rolling estimation, we can obtain samples of $x_{1:L+1}$ and $\theta$ sequentially, simply by skipping Step 2. 
The advantage of using our SMC-based method is that we can obtain the estimate of marginal likelihood $p(y_{1:L+1})$ as a by-product (the initializing algorithm and the marginal likelihood estimator are described in detail in the Supplementary Material B.).
This initializing algorithm can be used for the ordinary sequential learning of $\pi(x_{1:t},\theta \mid y_{1:t})\ (t = 1,\dots,T)$. We note that this approach is derived from the particle Gibbs scheme in \shortciteN{AndrieuDoucetHolenstein(10)}, and hence our approach is different from that of SMC$^2$ which applies the particle MH scheme as noted in \shortciteN{Chopin2013} and \shortciteN{Fulop2013}.  \\

\section{Illustrative examples}
\label{sec: Applications}
This section demonstrates the efficiencies of our proposed algorithm using two illustrative examples. The simple rolling-window sampler suffers from the serious weight degeneracy problem, while (the idealised and the practical) double-block samplers  overcome such difficulties. To evaluate the weight degeneracy in each of Steps 1 and 2, we define two ratios:
\begin{eqnarray}
R_{1t} = \frac{\rm{ESS}_{[s-1:t]}}{\rm{ESS}_{[s-1:t-1]}},
\quad
R_{2t} = \frac{\rm{ESS}_{[s:t]}}{\rm{ESS}_{[s-1:t]}}.
\end{eqnarray}
The ratio $R_{1t}$ measures the relative change of ESS in Step 1 after adding $y_t$ when compared with that of the previous step. If the distribution of particles is close to the posterior distribution from which we aim to sample in the step, $R_{1t}$  would be close to 1. On the other hand, in the presence of the weight degeneracy problem, it will be close to 0. Similarly, the ratio $R_{2t}$ measures the relative change of ESS in Step 2 after removing $y_{s-1}$ compared with that of the previous step.

\subsection{Linear Gaussian state space model}
\label{sec: Linear Gaussian}
\noindent
We first consider the following univariate linear Gaussian state space model:
\begin{eqnarray*}
y_t &=& x_t + \epsilon_t,\ \epsilon_t \sim \mathcal{N}(0, \sigma^2),\ t = 1,\dots,2000 \\
x_{t+1} &=& \mu + 0.25 (x_t - \mu) +  \eta_t,\ \eta_t \sim \mathcal{N}(0,2\sigma^2),\ t = 1,\dots,2000,\\
x_1 &=& \mu + \frac{\eta_0}{\sqrt{1-0.25^2}},\ \eta_0 \sim \mathcal{N}(0,2\sigma^2),
\end{eqnarray*}
where $\theta = (\mu,\sigma^2)'$ is a parameter vector. We adopt weak conjugate priors, $\mu \mid \sigma^2 \sim \mathcal{N}(0,10\sigma^2)$ and $\sigma^2 \sim \mathcal{IG}(5/2,0.05/2)$ where $\mathcal{IG}(a,b)$ denotes an inverse gamma distribution with shape parameter $a$ and scale parameter $b$. The rolling estimation is conducted with a window $[t-999,t]$,  $t=1001,\ldots,2000$ and $N=1000$ using the particle rolling MCMC with and without the double-block sampling. We choose $K=1,2,3,5$ and $10$ to investigate the effect of the block size. Since an idealised double-block sampler is feasible in the linear Gaussian state space model,  we compare the following three samplers:
\begin{enumerate}
\item Simple rolling-window sampler (as a benchmark).
\item Idealised double-block sampler.
\item Practical double-block sampler with $M = 100,300$ and $500$.
\end{enumerate}
\noindent
Table \ref{table: num of resampling} shows the number of resampling steps for three samplers.
For the simple rolling-window sampler, the resampling steps are triggered 1027 times, while they are drastically reduced for the double-block samplers. They decrease as we increase $K$ where the magnitude of the reduction is largest at $K=2$. 
For $K=2$, they are around 0.8\% and 7.2\% of the simple rolling-window sampler for the idealised and practical double-block samplers respectively.
Additionally, the number of resampling steps of the practical double-block sampler decreases to that of the idealised double-block sampler as $M$ increases.

\begin{figure}[H]
\centering
\caption{
The histograms of $R_{1t}$ (left) and $R_{2t}$ (right) ($t=1001,\ldots,2000$) for the simple rolling-window sampler (dotted blue) and the practical double-block sampler with $K=2$ and $M=100$ (solid red). 
}\label{Fig_simple_degeneracy_comparison-1}
\includegraphics[width=10cm]{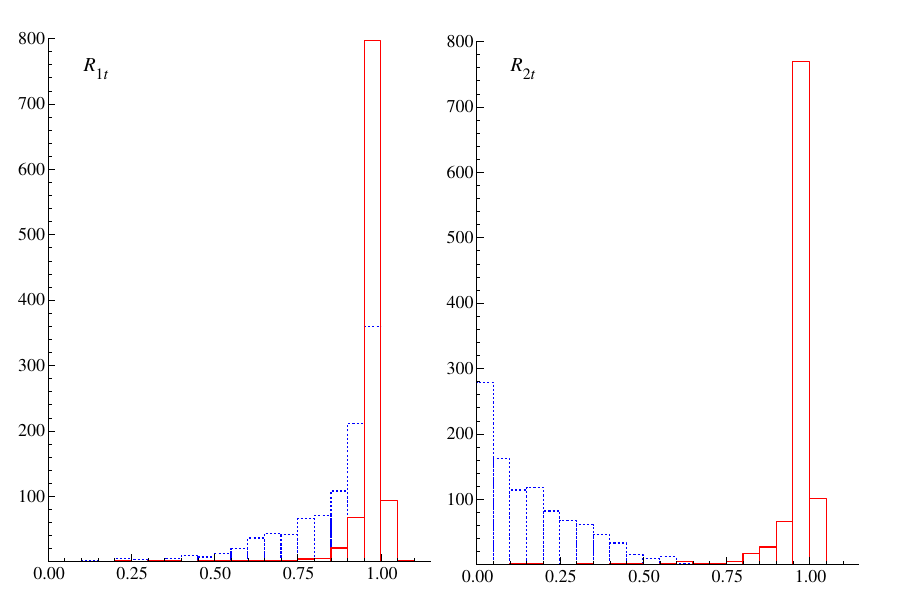}
\end{figure}

\begin{table}[H]
  \caption{The number of resampling steps of three samplers.}
  \label{table: num of resampling}
  \centering
  \begin{tabular}{lrrrrrr}
    \hline    \multicolumn{1}{c}{\small{Simple}} &   & \multicolumn{1}{c}{\small{Idealised}} & \multicolumn{3}{c}{Practical}\\ 
        \cline{4-6}
	     & $K$ &   & $M:100$ & $300$ & $500$ 
       \\ \hline              
	     & 1  & 48 &104 & 71 & 61 \\ 
	     & 2  & 8  & 74 & 33 & 23 \\ 
	 1027& 3  & 7  & 74 & 31 & 22 \\
             & 5  & 6  & 72 & 32 & 22 \\ 
             & 10 & 5  & 69 & 31 & 23 \\ \hline
  \end{tabular}

\end{table}
Figure \ref{Fig_simple_degeneracy_comparison-1} shows histograms of $R_{1t}$ and $R_{2t}$  for the simple rolling-window sampler and the practical double-block sampler with $K=2$ and $M=100$. The ratios $R_{1t}$  and $R_{2t}$ measure the relative magnitude of the effective sample size in Step 1 and Step 2  after adding $y_t$  and removing $y_{s-1}$ respectively when compared with that of the previous step at time $t$. 
The $R_{1t}$ values for the practical double-block sampler  are larger and less dispersed compared with those for the simple rolling-window sampler, suggesting that the forward block sampling is more efficient.
Additionally, the $R_{2t}$ values for the practical double-block sampler  are much larger and much less dispersed than those for the simple rolling-window sampler, which implies that the backward block sampling is much more efficient.  

\begin{figure}[H]
\centering
\caption{
The scatter plot of $R_{2t}$ versus $R_{1t}$ ($t=1001,\ldots,2000$) for the simple rolling-window sampler (blue plus) and the practical double-block sampler with $K=2$ and $M=100$ (red circle).  
}\label{Fig_simple_degeneracy_comparison-2}
\includegraphics[width=10cm]{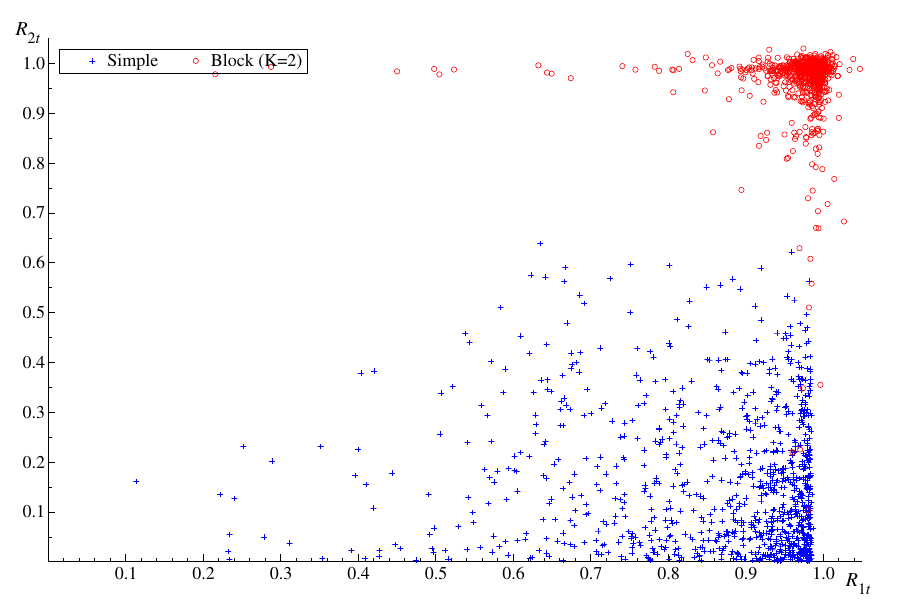}
\end{figure}

\begin{figure}[H]
\centering
\caption{
True posterior means and 95\% credible intervals (dotted black) with their estimates (solid red) for $\mu$ and $\sigma^2$ using the practical double-block sampler.}\label{Fig4}
\includegraphics[width=10cm]{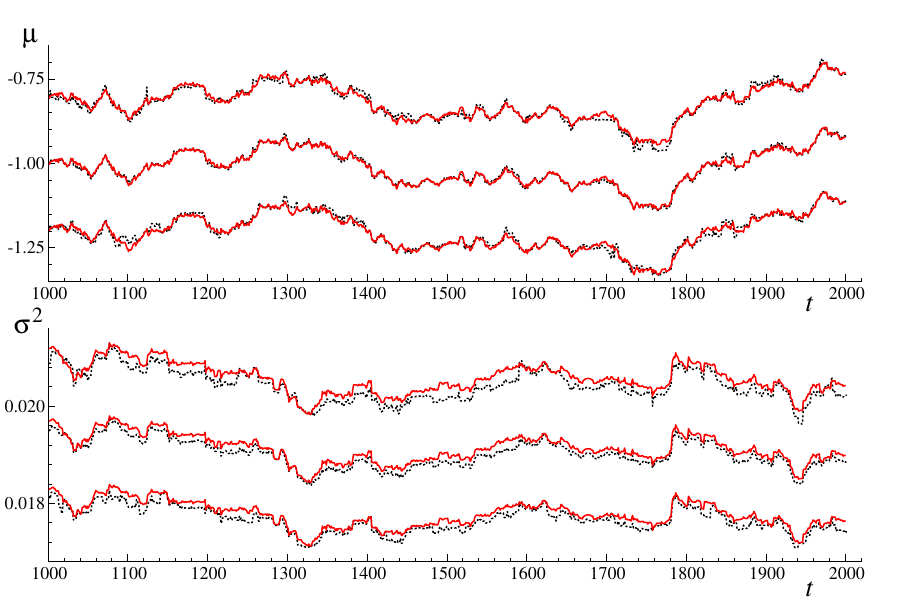}
\end{figure}

\begin{figure}[H]
\centering
\caption{Top: true log marginal likelihoods $\log p(y_{t-999:t})$ (dotted black) and their estimates (solid red) by the practical double block sampler.
Bottom: estimation errors $\log \hat{p}(y_{t-999:t}) - \log p(y_{t-999:t})$ for $t=1001,\ldots,2000$.
}\label{Fig5}
\includegraphics[width=10cm]{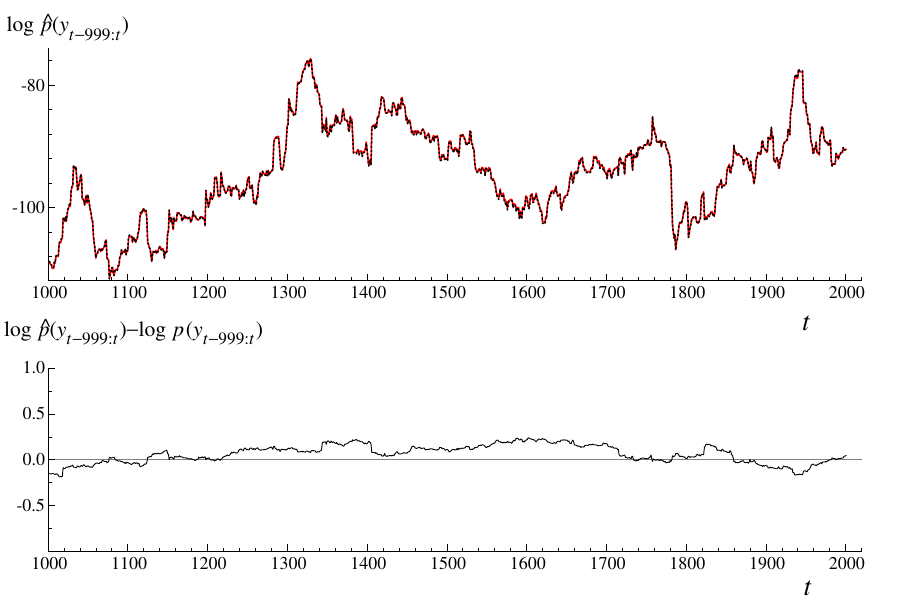}
\end{figure}

\noindent
Further, the scatter plots of $R_{1t}$ and $R_{2t}$ are shown in Figure \ref{Fig_simple_degeneracy_comparison-2} for two sampling methods. These results demonstrate that our practical double-block sampler is more efficient at both Steps 1 and 2 of each rolling step.
\noindent
Table \ref{table: stat of ESS} shows the summary statistics of $R_{1t}$ and $R_{2t}$.
The average of $R_{1t}$ for the practical double-block sampler is slightly larger than that for the simple rolling-window sampler, but the standard deviation for the former is less than half of that for the latter. Moreover, the average of $R_{2t}$ for the double-block sampling is six times larger than that for the simple sampling, while the standard deviation for the former is approximately half of that for the latter. Thus the practical double-block sampler drastically alleviate the weight degeneracy compared with the simple rolling-window sampler.
\begin{table}[H]
  \centering
  \caption{Summary statistics of $R_{1t}$ and $R_{2t}$ for the simple rolling-window sampler \vspace{-4mm}}
   and the practical double-block sampler ($K=2, M=100$)\vspace{2mm}\\
  \label{table: stat of ESS}
  \begin{tabular}{llcc}
    \hline
	                      & Method    &  Mean  & Std. dev. \\ \hline
	     $R_{1t}$         & Simple    & 0.862  & 0.145\\ 
	                      & Practical & {\bf 0.975}  & {\bf 0.057}\vspace{2mm}\\ 
	     $R_{2t}$         & Simple    & 0.161  & 0.139\\ 
	                      & Practical & {\bf 0.970}  & {\bf 0.068}\\ \hline
  \end{tabular}
\end{table}
\noindent 
Finally, to assess the accuracy of the practical double-block sampler (with $K=2$ and $M = 100$), we compare the estimation results with their corresponding analytical solutions. The particles are `refreshed' in the MCMC update step so that the approximation errors do not accumulate over time. In Figure \ref{Fig4}, the algorithm seems to correctly capture both means and 95\% credible intervals of the target posterior distribution. 
In Figure \ref{Fig5}, true log marginal likelihoods and their estimates are shown in with errors. The estimation errors are very small overall, implying that the proposed algorithm estimates the marginal likelihood $p(y_{t-999:t})$ accurately for $t=1001,\ldots,2000$.
\subsection{Realized stochastic volatility model}
This subsection considers the RSV model given by (\ref{eq:rsv-1})-(\ref{eq:rsv-5}) where the idealised double-block sampler is not feasible.  
For the static parameter $\theta=(\mu,\phi,\sigma^2_\eta,\xi,\sigma^2_u,\rho)'$, we assume the prior distributions as in \shortciteN{Takahashi2009}: 
\begin{align}
&    \frac{\phi + 1}{2} \sim \mathcal{B}(20,1.5),\ 
    c \sim \mathcal{N}(0,10),\ \sigma_u^2 \sim \mathcal{IG}(5/2,0.05/2),
\\
&    \Sigma \sim \mathcal{IW}(5,\Sigma_0), \ 
    \Sigma_0
    = 
    \left(
        5 \left[
            \begin{array}{cc}
                1 & -0.3 \times 0.1 \\
               -0.3 \times 0.1 & 0.01
            \end{array}
        \right]
    \right)^{-1}.
\end{align}
using the transformation
\begin{align}
    \sigma_\epsilon &= \exp(\mu / 2),\ 
    c  = \xi + \mu, \
    \Sigma =
    \left[
        \begin{array}{cc}
            \sigma^2_\epsilon & \rho \sigma_\epsilon \sigma_\eta \\
        \rho \sigma_\epsilon \sigma_\eta & \sigma^2_\eta
        \end{array}
    \right],
\end{align}
where $\mathcal{B}(a,b)$ and $\mathcal{IW}(r,S)$ denote a beta distribution with parameters $(a,b)$, and an inverse Wishart distribution with $r$ degrees-of-freedom and the scale matrix $S$ respectively.
\noindent
For $y_{1t}$ and $y_{2t}$, we use Standard and Poor's (S\&P) 500 index data, which are obtained from the Oxford-Man Institute Realized Library\footnote{
The data is downloaded at http:\slash\slash realized.oxford-man.ox.ac.uk\slash data\slash download
}
created by \shortciteN{Heber2009} (see \shortciteN{shephard2010realising} for details).
The initial estimation period is from January 1, 2000 $(t=1)$ to December 31, 2007 $(t=1988)$ with $L+1= 1988$.
The rolling estimation started after this initial sample period and moved the window until December 30, 2008 ($T=2248$). Thus the first estimation period is before the financial crisis caused by the bankruptcy of Lehman Brothers and the last estimation period includes the crisis.

\begin{figure}[H]
\centering
\caption{
Trace plot of estimated posterior means and 95\% credible intervals for parameters for S\&P500 return in RSV model (from December 31, 2007 to December 30, 2016) using the practical double-block sampler with $5$ (dotted blue) and 10 (solid red) iterations for the MCMC step.}\label{ASVforSP500} 
\includegraphics[width=15cm]{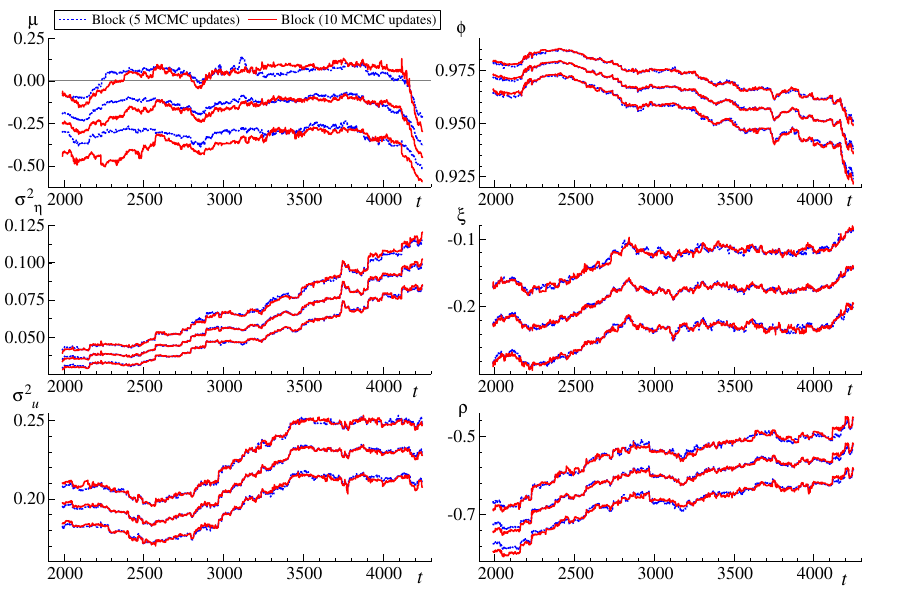}
\end{figure}

\begin{figure}[H]
\centering
\caption{
 Cumulative computation times (wall time, unit time = second) $(t=1988,\dots,4248)$.
}
\label{ESS_data_sp2}
\includegraphics[width=12cm]{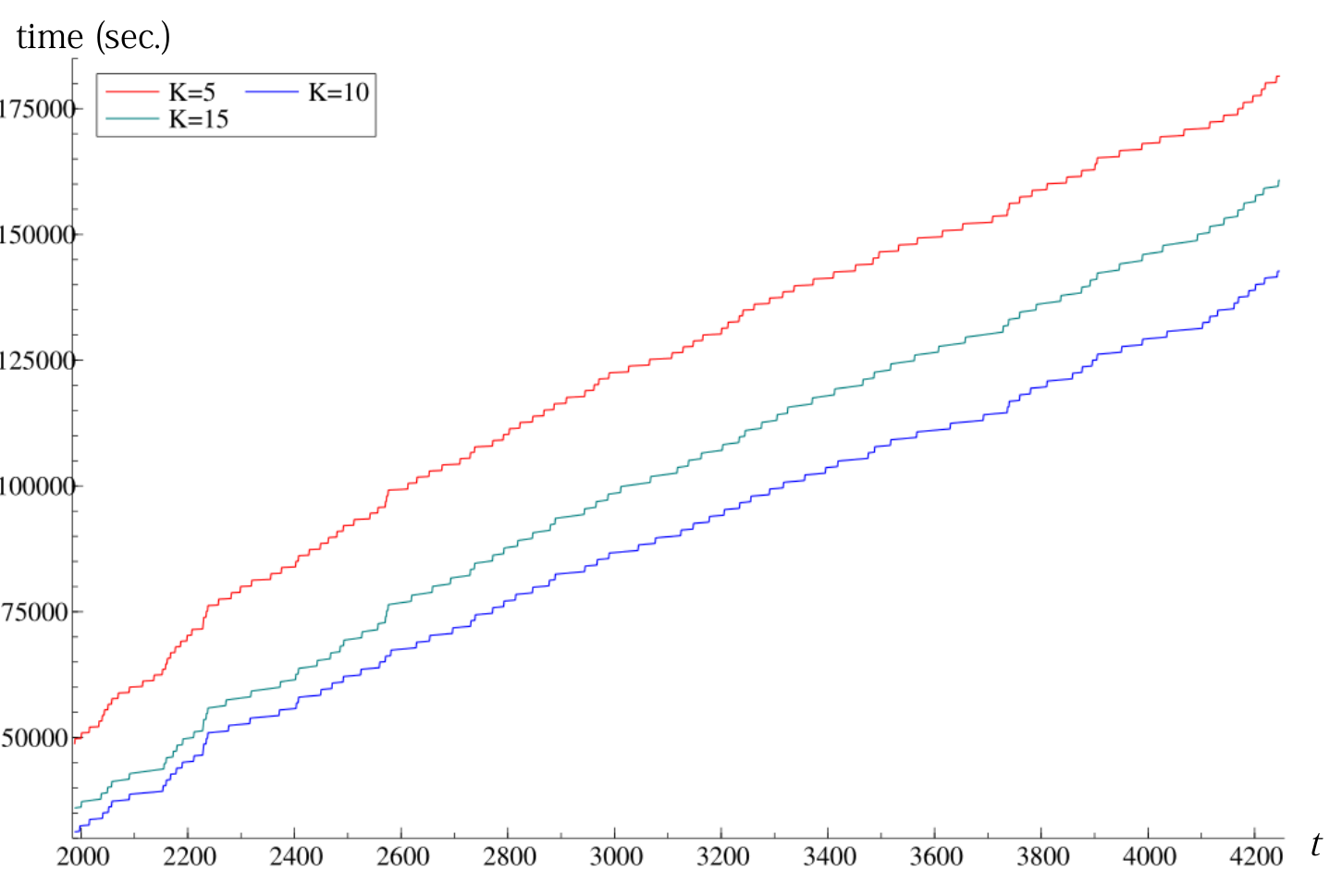}
\end{figure}

\begin{figure}[H]
\centering
\caption{The estimated posterior cumulative distribution functions of $\theta$.}
\vspace{-2mm}
MCMC (solid gray) and  practical double-block sampler: \\
1 (dashed green), 5 (dotted blue) and 10 (solid red) iterations.
\label{RSV1to1988}
\includegraphics[width=15cm]{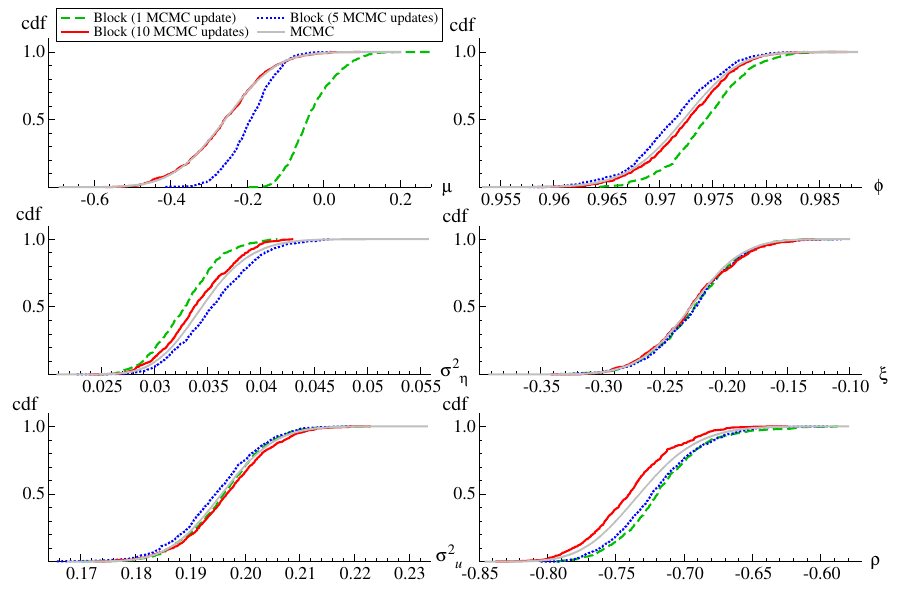}
\end{figure}

We first implement the simple rolling-window sampler. If the ESS is less than the threshold ($0.5\times N)$, the particles are refreshed with the MCMC update 10 times. (see  \shortciteN{Takahashi2009} for the details of the MCMC sampling). We set $N=1000$ and construct the proposal density $q_{t,\theta}(x_t \mid x_{t-1},y_{s-1:t})$ based on the normal mixture approximation (see \shortciteN{Omori2007}), which is expected to improve the weight degeneracy. 
%
Table \ref{table:ratio-rsv-simple}  presents a summary of $R_{1t}$ and $R_{2t}$.
As expected, $R_{2t}$'s are low, so the update with MCMC kernel should be implemented in almost every step. The results for $R_{1t}$'s also indicate that the ESS will be often less than the threshold to resample all the particles. In fact, due to these problems, the resampling steps are implemented 271 times for 260 data windows.
\begin{table}[H]
\centering
\caption{
Summary statistics for $R_{1t}$ and $R_{2t}$ ($t=1988,\dots,2248)$
\label{table:ratio-rsv-simple}}
\vspace{-3mm}
for the simple rolling-window sampler.
\\
\begin{tabular}{lccc}
 & Mean & Median & Std. dev. \\ \hline
$R_{1t}$ & 0.837 & 0.912 & 0.193 \\ 
$R_{2t}$ & 0.227 & 0.197 & 0.176\\
\hline
\end{tabular}
\normalsize
\end{table}
\noindent
Next, we implement the practical double-block sampler with\footnote{We also tried using other values of $M$ but the computation time is the shortest with $M= 300$.} $M=300$ and $N=1000$. Further we always implement 10 MCMC iterations below unless otherwise stated. As a proposal density, we simply use a prior density $q_{t,\theta}(x_t \mid x_{t-1},y_{s-1:t}) = f_{\theta}(x_t \mid x_{t-1})$ to demonstrate that the practical double-block sampler improves even when using the simple proposal.
%
%
%
The summary statistics of $R_{1t}$ and $R_{2t}$ are shown in Table \ref{table:ratio-rsv} where we use  $K=5,10$ and $15$. 
In contrast to the simple rolling-window algorithm, both means are close to 1 demonstrating that our proposed algorithm succeeded in  overcoming the weight degeneracy problem. 
 As $K$ increases, $R_{1t}$ and $R_{2t}$ become larger and less dispersed, but the difference becomes smaller for $K=10$ and $K=15$.
\begin{table}[H]
\centering
\caption{Summary statistics for $R_{1t}$ and $R_{2t}$ $(t=1988,\dots,4248)$ \label{table:ratio-rsv}}
\vspace{-3mm}
using the practical double-block sampler.
\vspace{1mm}\\
  \begin{tabular}{llccc} \hline
             & $K$& Mean & Median & Std. dev.\\ \hline
    $R_{1t}$ & 5  & 0.981 & 0.995 & 0.058 \\
             & 10 & 0.985 & 0.996 & 0.053 \\
             & 15 & 0.986 & 0.997 & 0.055 \vspace{2mm}\\
    $R_{2t}$ & 5  & 0.983 & 0.993 & 0.044 \\
             & 10 & 0.988 & 0.994 & 0.036 \\
             & 15 & 0.988 & 0.994 & 0.035 \\ \hline
  \end{tabular}
\normalsize
\end{table}
\noindent
Figure \ref{ASVforSP500} shows the trace plot of estimated posterior means and 95\% credible intervals for $\theta=(\mu,\phi,\sigma^2_\eta,\xi,\sigma^2_u,\rho)'$ from December 31, 2007 $(t=1988)$ to December 30, 2016 $(t=4248)$. 
From the rolling estimation results, we are able to observe the transition of the economic structure and the effect of the financial crisis ( $t=2150,\ldots,2213$ correspond to September, October and November in 2008) .
The posterior distribution of $\mu$ seems to be stable before $t=4000$ (January 7, 2016), but its mean and 95\% intervals decrease after $t=4000$. The average level of log volatility started to decrease sharply toward the end of the sample period. The autoregressive parameter, $\phi$, continues to decrease throughout the sample period indicating that the latent log volatility becomes less persistent. The variances, $\sigma_{\eta}^2$ and $\sigma_u^2$, of error terms in the state equation and the measurement equation of the log realized volatility continue to increase, while the bias adjustment term, $\xi$, and the leverage effect, $\rho$, become closer to zero during the sample period. The leverage effects in the stock market are weaker after the financial crisis. 
\noindent
Figure \ref{ESS_data_sp2} shows three cumulative computation times (wall time) for the same period corresponding to $K=5,10$ and $15$. The computation times with $K=5$ and $K = 15$ are longer than that with $K=10$. This finding implies that, when $K=5$, the effect of the blocking is not sufficient to reduce the path dependence between $x_t$ and $x_{t-K-1}$ (similarly,  $x_{s-1}$ and $x_{s+K}$). When $K = 15$, the Monte Carlo error in the local conditional SMC increased the variance of the importance weights even though there is a certain decrease in the variance due to the increase in $K$ (we shall see more details in Section 5)\footnote{Also see Supplementary Material C 
for the comparison of the computation time of the practical double-block sampler with those of the MCMC and the particle MCMC.}.
\noindent
Finally, in Figure \ref{RSV1to1988}, we investigate the effect of the number of iterations in the MCMC steps on the estimation accuracy of the posterior distribution function of $\theta$ for the proposed sampling algorithm. 
The estimation period is from January 1, 2000 to December 31, 2007 ($t=1,\ldots,1988$). First, the MCMC sampling is conducted to obtain the accurate estimates of the distribution functions (solid gray). Then we apply our practical double-block sampler with $K = 10$, $M=300$ and $N=1000$ for three cases: one, five and ten MCMC updates. 
Among three cases, the estimates obtained by  5 or 10 iterations are close to those obtained by the exact MCMC sampling. If  only one iteration is performed in the MCMC update step, the estimation results are found to be inaccurate because the MCMC iterations not only diversify the particles but also correct approximation errors introduced by the particle algorithm, which basically update only a part of the vector $x^n_{s-1:t}$. The estimation errors for the distribution function of $\mu$ are most serious, probably because the mixing property of MCMC sampling in the RSV model is poor especially with respect to $\mu$ as discussed in the numerical studies of \shortciteN{Takahashi2009}. Thus these results suggest that MCMC iterations should be implemented a sufficient number of times in the MCMC update steps such that the particles can trace the correct posterior distributions.
\section{Theoretical justification}
In this section, we provide theoretical justifications of our proposed algorithm in Section \ref{sec: Marginalized block sampling tools}. We prove that our posterior density is obtained as a marginal density of the artificial target density.

\subsection{Forward block sampling}
\label{sec: Justification: block sampling in augmented space}
\noindent
{\it The artificial target density and its marginal density.}
We prove that our posterior density of $(x_{s-1:t}^n,\theta^n)$ given $y_{s-1:t}$ is obtained as a marginal density of the artificial target density in the forward block sampling. The superscript $n$ will be suppressed for simplicity below.

In Step 1-1(a) of Section \ref{sec: MBS to incorporate the new information}, the probability density function of $(x^{k_{t-K}}_{t-K},\dots,x^{k_{t-1}}_{t-1})=x_{t-K:t-1}$ and $(a^{k_{t-K+1}}_{t-K},\dots,a^{k_{t}}_{t-1})$ given $(x_{t-K-1},\theta)$ and $y_{t-K:t-1}$ is
\begin{eqnarray}
p(x_{t-K:t-1},a^{k_{t-K+1}}_{t-K},\dots,a^{k_{t}}_{t-1}\mid x_{t-K-1}, y_{t-K:t-1},\theta)
&=&\frac{\pi(x_{t-K:t-1}\mid x_{t-K-1}, y_{t-K:t-1},\theta)}{M^{K}}.\nonumber
\\
&&
\end{eqnarray}

\noindent 
Let $a_{j}^{1:M}=(a_j^1,\ldots,a_j^M)$, $x_{j}^{1:M}=(x_{j}^{1},\ldots,x_{j}^{M})$ and $a^{-k_{j+1}}_{j} \equiv a^{1:M}_{j} \setminus a_j^{k_{j+1}}= a^{1:M}_{j} \setminus k_{j}$ for $j=t-K,\ldots, t-1$ where we note $a^{k_{j+1}}_{j}=k_j$ and $k_{t}=1$ in (\ref{rule on indices}).
Further, let $a_{t-K:t-1}^{1:M}=\{a_{t-K}^{1:M},\ldots, a_{t-1}^{1:M}\}$, and $x_j^{-k_j}=\{x_j^{a_j^1},\ldots,x_j^{a_j^M}\}\setminus x_j^{k_{j}}$. Then, in  1(b), 1(c) and 1(d) of Step 1, given $x_{t-K-1}$, 
$(x^{k_{t-K}}_{t-K},\dots,x^{k_{t-1}}_{t-1}) =  x_{t-K:t-1}$ and $(a^{k_{t-K+1}}_{t-K},\dots,a^{k_{t}}_{t-1})  =  (k_{t-K},\dots,k_{t-1})$, the probability density function of all variables is defined as 
\begin{eqnarray}
\small
\lefteqn{
\hspace{-5mm}
\psi_\theta \left(x_{t-K}^{-k_{t-K}},\ldots,x_{t-1}^{-k_{t-1}},x^{1:M}_{t},a_{t-K}^{-k_{t-K+1}},\ldots,a_{t-1}^{-k_{t}},k^*_t \mid x_{t-K-1:t-1}, a_{t-K}^{k_{t-K+1}},\ldots,a_{t-1}^{k_{t}}, y_{t-K:t}\right) } && \nonumber \\
& = & 
\prod^M_{\substack{m=1\\ m\neq k_{t-K}}}q_{t-K,\theta}(x_{t-K}^m \mid x_{t-K-1},y_{t-K})
\times \prod^{t-1}_{j=t-K+1} \prod^M_{\substack{m=1\\ m\neq k_j}} V_{j-1,\theta}^{a^m_{j-1}}q_{j,\theta}(x_j^m \mid x_{j-1}^{a^m_{j-1}},y_{j})\nonumber
\\
&  & 
\times 
q_{t,\theta}(x_t^1 \mid x_{t-1}^{k_{t-1}},y_{t}) 
\times
\prod^M_{m=2} V_{t-1,\theta}^{a^m_{t-1}}q_{t,\theta}(x_t^m \mid x_{t-1}^{a^m_{t-1}},y_{t})
\times
V^{k^*_t}_{t,\theta}.
\label{Psi in the filtering}
\end{eqnarray}

\noindent
In Step 1-2, we  multiply $W_{[s-1,t-1]}$ by $\hat{p}(y_t \mid x_{t-K-1}^n,y_{s-1:t-1},\theta^n)$ to adjust the importance weight for $W_{[s-1,t]}$.
Let $x_{t-K:t}^{1:M}=(x_{t-K}^{1:M},\ldots,x_{t}^{1:M})$ and $a_{t-K:t-1}^{1:M}=(a_{t-K}^{1:M},\ldots,a_{t-1}^{1:M})$.
Our artificial target density (before the particle smoother step) is written as
\begin{eqnarray}
\lefteqn{\hat{\pi}(x_{s-1:t-K-1},x^{1:M}_{t-K:t},a^{1:M}_{t-K:t-1},k^*_t,\theta \mid y_{s-1:t})}&&  \nonumber \\
&\equiv &
\frac{\pi(x_{s-1:t-K-1}, x_{t-K}^{k_{t-K}},\ldots,x_{t-1}^{k_{t-1}},\theta \mid y_{s-1:t-1})}{M^K}\nonumber\\
&&
\times 
\scalebox{0.8}{$\textstyle
\psi_\theta (x_{t-K}^{-k_{t-K}},\ldots,x_{t-1}^{-k_{t-1}},x^{1:M}_{t},a_{t-K}^{-k_{t-K+1}},\ldots,a_{t-1}^{-k_{t}},k^*_t\mid x_{t-K-1:t-1}, a_{t-K}^{k_{t-K+1}},\ldots,a_{t-1}^{k_{t}}, y_{t-K:t})$}
\nonumber\\
&& \times\frac{\hat{p}(y_t \mid x_{t-K-1},y_{s-1:t-1},\theta)}{p(y_t \mid y_{s-1:t-1})} \nonumber \\
&= & \frac{\pi(x_{s-1:t-1}, \theta \mid y_{s-1:t-1})}{M^{K}}\nonumber \\
&& \times 
\prod^M_{\substack{m=1\\ m\neq k_{t-K}}}q_{t-K,\theta}(x_{t-K}^m \mid x_{t-K-1},y_{t-K})
\times 
\prod^{t-1}_{j=t-K+1} \prod^M_{\substack{m=1 \\ m\neq k_j}} V_{j-1,\theta}^{a^m_{j-1}}q_{j,\theta}(x_j^m \mid x_{j-1}^{a^m_{j-1}},y_{j})
\nonumber \\
&& \times \quad 
q_{t,\theta}(x_t^1 \mid x_{t-1}^{k_{t-1}},y_{t})
\times 
\prod^M_{m=2} V_{t-1,\theta}^{a^m_{t-1}}q_{t,\theta}(x_t^m \mid x_{t-1}^{a^m_{t-1}},y_{t}) \times V^{k^*_t}_{t,\theta} \nonumber
\\
&&
\times
\frac{\hat{p}(y_t \mid x_{t-K-1},y_{s-1:t-1},\theta)}{p(y_t \mid y_{s-1:t-1})}.
\label{target dist in Block sampling in augmented space}
\end{eqnarray}
Note that $p(y_t \mid y_{s-1:t-1})$ is the normalizing constant of this target density, which will be shown in Proposition \ref{decomp of variance of incremental weight}.
The proposed forward block sampling is justified by proving that the marginal density of $(x_{s-1},\ldots,x_{t-K-1}, x_{t-K}^{k_{t-K}^*},\ldots,x_{t}^{k_t^*},\theta)$ in the above artificial target density $\hat{\pi}$ is $\pi(x_{s-1},\ldots,x_{t-K-1}, \\ x_{t-K}^{k_{t-K}^*},\ldots,x_{t}^{k_t^*},\theta \mid y_{s-1:t})$.
%
%
%
%
%
\begin{prop}
\label{prop 2}
The artificial target density $\hat{\pi}$ for the forward block sampling can be written as 
\begin{eqnarray}
\lefteqn{\hat{\pi}(x_{s-1:t-K-1},x^{1:M}_{t-K:t},a^{1:M}_{t-K:t-1},k^*_t,\theta \mid y_{s-1:t})}&&
\nonumber\\
&=&  
\frac{\pi(x_{s-1:t-K-1},x^{k_{t-K}^*}_{t-K},\ldots,x^{k_t^*}_{t},\theta \mid y_{s-1:t})}{M^{K+1}}
\times
\prod^M_{\substack{m=1\\ m\neq k^*_{t-K}}}q_{t-K,\theta}(x^m_{t-K} \mid x_{t-K-1},y_{t-K})
\nonumber\\
&& 
\times
\prod^t_{j=t-K+1}\prod^M_{\substack{m=1 \\ m\neq k^*_j}}V_{j-1,\theta}^{a^m_{j-1}}q_{j,\theta}(x_j^m\mid x_{j-1}^{a^m_{j-1}},y_{j}),
\label{true target dist in sequential est}
\end{eqnarray}
and 
the marginal density of $(x_{s-1:t-K-1}, x_{t-K}^{k_{t-K}^*},\ldots,x_{t}^{k_t^*},\theta)$ is $\pi(x_{s-1:t-K-1}, x_{t-K}^{k_{t-K}^*},\ldots,x_{t}^{k_t^*},\theta \mid y_{s-1:t})$.
\end{prop}
\noindent
{\it Proof}. See Supplementary Material A. \\

\noindent
Proposition \ref{prop 2} implies that we can obtain a posterior random sample $(x_{s-1:t},\theta)$ given $y_{s-1:t}$ (with the importance weight $W_{[s-1,t]}$) by sampling from the artificial target distribution $\hat{\pi}$. This justifies our proposed forward block sampling scheme.\\


\noindent
{\it Properties of the incremental weight}.
We consider the mean and variance of the (unnormalized) incremental weight, $\hat{p} (y_t \mid y_{s-1:t-1},x_{t-K-1},\theta)$. Proposition \ref{decomp of variance of incremental weight} shows that this weight can be considered an unbiased estimator.
%
%
%
%
%
\begin{prop}
\label{decomp of variance of incremental weight}
If 
\begin{eqnarray*}
(x_{s-1:t-K-1}, x_{t-K}^{k_{t-K}},\ldots,x_{t-1}^{k_{t-1}},k_{t-K:t-1},\theta) \sim \frac{\pi(x_{s-1:t-K-1}, x_{t-K}^{k_{t-K}},\ldots,x_{t-1}^{k_{t-1}},\theta \mid y_{s-1:t-1})}{M^K}
\end{eqnarray*}
 and
\begin{eqnarray*}
(x_{t-K}^{-k_{t-K}},\ldots,x_{t}^{-k_{t-1}},x^{1:M}_{t},a_{t-K}^{-k_{t-K+1}},\ldots,a_{t-1}^{-k_t},k^*_t)
 &\sim &\psi_\theta 
\end{eqnarray*}
where $\psi_\theta$ is given in (\ref{Psi in the filtering}), then
\begin{eqnarray*}
E[\hat{p}(y_t \mid x_{t-K-1},y_{s-1:t-1},\theta)|y_{s-1:t}] 
&=& E[p(y_t \mid x_{t-K-1},y_{s-1:t-1}, \theta)|x_{t-K-1},y_{s-1:t}, \theta]
\\
&=& p(y_t \mid y_{s-1:t-1}).
\end{eqnarray*}
\end{prop}
\noindent
{\it Proof}. See Supplementary Material A. \\

\noindent
This shows that the incremental weight $\hat{p}(y_t \mid x_{t-K-1},y_{s-1:t-1},\theta)$ is an unbiased estimator of  the conditional likelihood $p(y_t \mid x_{t-K-1},y_{s-1:t-1}, \theta)$ given $(x_{t-K-1},\theta)$. It is also an unbiased estimator of the marginal likelihood $p(y_t \mid y_{s-1:t-1})$ unconditionally, which implies that $p(y_t \mid y_{s-1:t-1})$ is a normalizing constant for the artificial target density $\hat{\pi}$. 

Further, from the law of total variance, we obtain the decomposition of the variance as follows. 
\begin{eqnarray*}
\lefteqn{
\mathrm{Var}[\hat{p}(y_t \mid x_{t-K-1},y_{s-1:t-1},\theta)\mid y_{s-1:t}]
} && \\
& = &  \mathrm{Var}[p(y_t \mid x_{t-K-1},y_{s-1:t-1},\theta)\mid y_{s-1:t}]
\\
&& \hspace{3cm}
+ E\left[\mathrm{Var}[\hat{p}(y_t \mid x_{t-K-1},y_{s-1:t-1},\theta) \mid x_{s-1:t-K-1},y_{s-1:t}, \theta]\right].
\end{eqnarray*}
\noindent
The variance of the incremental weight consists of two components, including variance of the conditional likelihood and (expected) variance which is introduced using $M$ particles to approximate the conditional likelihood. This decomposition identifies factors that influences the ESS of the particles.
Regarding the first component, for any positive integers, $K_1,K_2$, with $K_1<K_2$, the following inequality holds:
\begin{eqnarray*}
\mathrm{Var}[p(y_t \mid x_{t-K_1-1},y_{s-1:t-1},\theta)] \geq 
\mathrm{Var}[p(y_t \mid x_{t-K_2-1},y_{s-1:t-1},\theta)], 
\end{eqnarray*}
which is a straightforward result from the law of total variance for $p(y_t \mid x_{t-K_1-1},y_{s-1:t-1},\theta)$ using
\begin{eqnarray}
E\left[
p(y_t \mid x_{t-K_1-1},y_{s-1:t-1},\theta) \mid x_{s-1:t-K_2-1},\theta
\right]
=  p(y_t \mid x_{t-K_2-1},y_{s-1:t-1},\theta).
\end{eqnarray}
On the other hand, the second component may become large as $K$ increases, but it is expected to be controlled by changing the number of particles $M$.


\subsection{Backward block sampling}
\label{sec: Justification: backward block sampling in augmented space}
\noindent
{\it The artificial target density and its marginal density.}
This subsection proves that our posterior density of $(x_{s:t}^n,\theta^n)$ given $y_{s:t}$ is obtained as a marginal density of the artificial target density in the backward block sampling. The superscript $n$ will be suppressed for simplicity below.

In Step 2-1(a) of Section \ref{sec: Algorithm: inverted marginal block sampling scheme}, the probability density function of 
$(x^{k_{s-1}}_{s-1},\dots,x^{k_{s+K-1}}_{s+K-1})=x_{s-1:s+K-1}$ and $(a^{k_{s-2}}_{s-1},\dots,a^{k_{s+K-2}}_{s+K-1})$ given $(x_{s+K},\theta)$ and $y_{s-1:t}$ is
\begin{eqnarray}
p(x_{s-1:s+K-1},a^{k_{s-2}}_{s-1},\dots,a^{k_{s+K-2}}_{s+K-1}\mid x_{s+K}, y_{s-1:t},\theta)
&=&\frac{\pi(x_{s-1:s+K-1}\mid x_{s+K}, y_{s-1:t},\theta)}{M^{K+1}}.\quad \mbox{}
\end{eqnarray}
In 1(b), 1(c) and 1(d) of Steps 2, given $x_{s+K}$, 
$(x^{k_{s-1}}_{s-1},\dots,x^{k_{s+K-1}}_{s+K-1}) =  x_{s-1:s+K-1}$, $(a^{k_{s-2}}_{s-1},\dots,a^{k_{s+K-2}}_{s+K-1})  =  (k_{s-1},\dots,k_{s+K-1})$ and $y_{s-1:s+K-1}$, the probability density function of all variables is defined as 
\small
\begin{eqnarray}
\lefteqn{\bar{\psi}_\theta (x_{s-1}^{-k_{s-1}},\ldots,x_{s+K-1}^{-k_{s+K-1}},a_{s}^{-k_{s-1}},\ldots,a_{s+K-1}^{-k_{s+K-2}},k^*_{s} \mid x_{s-1:s+K}, a_{s-1}^{k_{s-2}}, \ldots,a_{s+K-1}^{k_{s+K-2}},y_{s-1:s+K-1})} & & \nonumber \\
& = & \prod^M_{\substack{m=1 \\ m\neq k_{s+K-1}}} q_{s+K-1,\theta}(x^m_{s+K-1} \mid x_{s+K},y_{s+K-1})
\times
\prod^{s+K-2}_{j=s-1} \prod^M_{\substack{m=1 \\m\neq k_j}} V^{a^m_{j+1}}_{j+1,\theta}q_{j,\theta}(x^m_j \mid x^{a^m_{j+1}}_{j+1},y_{j})
\times V^{k^*_{s}}_{s,\theta}.\nonumber\\
&&
\label{Psi in the forgetting}
\end{eqnarray}
\normalsize

\noindent
In Step 2-2, we  divide $W_{[s-1,t]}$ by $\hat{p}(y_{s-1} \mid x_{s+K}^n,y_{s:t},\theta^n)$ to adjust the importance weight for $W_{[s,t]}$.
%
Similarly to the discussion in Section \ref{sec: Justification: block sampling in augmented space}, we consider an extended space with the artificial target density written as 
\begin{eqnarray}
\lefteqn{\check{\pi}(x_{s-1:s+K-1}^{1:M},x_{s+K:t},a_{s:s+K-1}^{1:M},k_{s-1},k^*_s,\theta \mid y_{s-1:t})}&&  \nonumber \\
&\equiv &
\frac{\pi(x_{s-1:t}, \theta \mid y_{s-1:t})}{M^{K+1}}\nonumber \\
& \times & \prod^M_{\substack{m=1 \\ m\neq k_{s+K-1}}} q_{s+K-1,\theta}(x^m_{s+K-1} \mid x_{s+K},y_{s+K-1})
\times
\prod^{s+K-2}_{j=s-1} \prod^M_{\substack{m=1 \\m\neq k_j}} V^{a^m_{j+1}}_{j+1,\theta}q_{j,\theta}(x^m_j \mid x^{a^m_{j+1}}_{j+1},y_{j})
\nonumber\\
&\times& V^{k^*_{s}}_{s,\theta}
\times
\frac{p(y_{s-1} \mid y_{s:t})}{\hat{p}(y_{s-1} \mid x_{s+K},y_{s:t},\theta)},
\label{target dist in rolling est}
\end{eqnarray}
where $p(y_{s-1} \mid y_{s:t})^{-1}$ is the normalizing constant of this target density as shown in Proposition \ref{decomp of variance of incremental weight in discarding steps}. 
Below we state Proposition \ref{prop in discarding steps} for the backward block sampling, which correspond to Proposition \ref{prop 2} for the forward block sampling.
%
%
%
%
%
%
%
%
\begin{prop}
\label{prop in discarding steps}
The artificial target density $\check{\pi}$ for the backward block sampling can be rewritten as
\begin{eqnarray}
\lefteqn{
\check{\pi}(x_{s-1:s+K-1}^{1:M},x_{s+K:t},a_{s:s+K-1}^{1:M},k_{s-1},k^*_{s},\theta \mid y_{s-1:t})
} && \nonumber\\
&& = \frac{\pi(x_{s}^{k_{s}^*},\ldots,x_{s+K-1}^{k_{s+K-1}^*},x_{s+K:t},\theta \mid y_{s:t})}{M^{K}}
\times
\prod^M_{\substack{m=1\\ m\neq k^*_{s+K-1}}} q_{s+K-1,\theta}(x^m_{s+K-1} \mid x_{s+K},y_{s+K-1:t})
  \nonumber\\
&&
\times
\prod^{s+K-2}_{j=s} \prod^M_{\substack{m=1\\ m\neq k^*_j}} V^{a^m_{j+1}}_{j+1,\theta}q_{j,\theta}(x^m_j \mid x^{a^m_{j+1}}_{j+1},y_{j})
\times 
\prod^M_{m=1}V^{a^m_{s}}_{s,\theta}q_{s-1,\theta}(x^m_{s-1}\mid x^{a^m_{s}}_{s},y_{s-1})
\times
V^{k_{s-1}}_{s-1,\theta}, \hspace{8mm}\mbox{}
\label{true target dist in rolling est}
\end{eqnarray}
and the marginal density of $(x_{s}^{k_{s}^*},\ldots,\allowbreak x_{s+K-1}^{k_{s+K-1}^*}, x_{s+K:t},\theta)$ is 
 $\pi(x_{s}^{k_{s}^*},\ldots, x_{s+K-1}^{k_{s+K-1}^*}, x_{s+K:t}, \theta \mid y_{s:t})$. 
\end{prop}
\noindent
{\it Proof}. See Supplementary Material A. \\

\noindent
Although the probability density (\ref{true target dist in rolling est}) in Proposition \ref{prop in discarding steps}  has a bit different form from that of (\ref{true target dist in sequential est}) in Proposition \ref{prop 2}, its marginal probability density is found to be the target posterior density $\pi(x_{s:t}, \theta \mid y_{s:t})$.

\noindent
{\it Properties of the incremental weight}.
Similar results to Proposition \ref{decomp of variance of incremental weight} hold for the backward block sampling, and are summarized in Proposition \ref{decomp of variance of incremental weight in discarding steps}.
\begin{prop}
\label{decomp of variance of incremental weight in discarding steps}
If 
\begin{eqnarray*}
(x_{s-1}^{k_{s-1}},\ldots,x_{s+K-1}^{k_{s+K-1}}, x_{s+K:t},k_{s-1:s+K-1},\theta) \sim \frac{\pi(x_{s-1}^{k_{s-1}},\ldots,x_{s+K-1}^{k_{s+K-1}}, x_{s+K:t},\theta \mid y_{s-1:t})}{M^{K+1}}
\end{eqnarray*}
and
\begin{eqnarray*}
(x_{s-1}^{-k_{s-1}},\ldots,x_{s+K-1}^{-k_{s+K-1}},a_{s}^{-k_{s-1}},\ldots,a_{s+K-1}^{-k_{s+K-2}},k^*_{s})
 &\sim &\bar{\psi}_\theta,
\end{eqnarray*}
where $\bar{\psi}_\theta$ is given in (\ref{Psi in the forgetting}), then
\begin{eqnarray*}
E[\hat{p}(y_{s-1} \mid x_{s+K},y_{s:t},\theta)^{-1}]
& = & 
E[\hat{p}(y_{s-1} \mid x_{s+K},y_{s:t},\theta)^{-1}\mid x_{s+K},y_{s:t},\theta]
\\
&=& p(y_{s-1} \mid y_{s:t},\theta)^{-1}.
\end{eqnarray*}
\end{prop}
\noindent
{\it Proof}. See Supplementary Material A. 

\subsection{Particle simulation smoother}
In \shortciteN{WhiteleyAndrieuDoucet(10)} and the discussion of Whiteley following \shortciteN{AndrieuDoucetHolenstein(10)}, the additional step is introduced to explore all possible ancestral lineages. This is expected to improve the mixing property of the particle Gibbs (see e.g. \shortciteN{Chopin2015}, \shortciteN{LeeSinghVihola(20)}).  We also incorporate such a particle simulation smoother into the double block sampling based on the following proposition. 

\begin{prop}
\label{prop:particle_ss}
The joint conditional density of  $(k_{t-K}^*,\ldots,k^*_{t})$  is given by
\begin{eqnarray}
\lefteqn{\hat{\pi}(k_{t-K}^*,\ldots,k^*_{t}|x_{s-1:t-K-1},x_{t-K:t}^{1:M},a_{t-K:t-1}^{1:M},y_{s-1:t},\theta)} && \nonumber \\
& = & 
\hat{\pi}(k^*_{t}|x_{s-1:t-K-1},x_{t-K:t}^{1:M},a_{t-K:t-1}^{1:M},y_{s-1:t},\theta) 
\nonumber \\
&\times&
\prod_{t_0=t-1}^{t-K}\hat{\pi}(k_{t_0}^*|x_{s-1:t-K-1},x_{t-K:t_0}^{1:M},a_{t-K:t_0-1}^{1:M},x_{t_0+1}^{k_{t_0+1}^*},\ldots,x_{t}^{k_{t}^*},k^*_{t_0+1:t}, y_{s-1:t},\theta), 
\end{eqnarray}
where
\begin{eqnarray}
\lefteqn{
\hat{\pi}(k_{t_0}^*|x_{s-1:t-K-1},x_{t-K:t_0}^{1:M},a_{t-K:t_0-1}^{1:M},x_{t_0+1}^{k_{t_0+1}^*},\ldots,x_{t}^{k_{t}^*},k^*_{t_0+1:t}, y_{s-1:t},\theta)} && 
\nonumber \\
&&\quad =
\bar{V}^{k^*_{t_0}}_{t_0,\theta},
\quad 
\bar{V}^{m}_{j,\theta}
\equiv
\frac{V^{m}_{j,\theta}f_\theta(x^{k^*_{j+1}}_{j+1} \mid x^{m}_{j},y_{j+1})}
{\sum_{i=1}^MV^{i}_{j,\theta}f_\theta(x^{k^*_{j+1}}_{j+1} \mid x^{i}_{j},y_{j+1})}.\hspace{5cm}\mbox{}
\label{inportance weight in BS}
\end{eqnarray}
\end{prop}
\noindent
{\it Proof}. See Supplementary Material A. \\

\noindent
Suppose we have $(x_{s-1:t-K-1},x_{t-K:t}^{1:M},a_{t-K:t-1}^{1:M},k^*_t,\theta) \sim \hat{\pi}$ where $\hat{\pi}$ is defined in (\ref{target dist in Block sampling in augmented space}).
In Step 1-1(d), the lineage $k^*_{t-K:t}$ is automatically determined when $k^*_t$ is chosen. The particle simulation smoother breaks this relationship and again samples $k^*_{t-K:t}$ jointly by generating $k^*_{j}\sim \mathcal{M}(\bar{V}_{j,\theta}^{1:M}),$ $j=t-1,\dots,t-K$, recursively.


%
%
\section{Conclusion}
In this paper, we propose a novel efficient estimation method to implement the rolling-window particle MCMC simulation using a SMC framework and refreshing steps with MCMC kernel.
The weighted particles are updated to learn and discard the information of the new and old observations using the forward and backward block sampling based on the conditional SMC algorithm, which effectively circumvent the weight degeneracy problem. The proposed estimation methodology is also applicable to the ordinary sequential estimation with parameter uncertainty.  The illustrative examples show that our proposed sampler outperforms the simple rolling-window sampler.
\small
\section*{Acknowledgement}
All computational results in this paper are generated using Ox metrics 7.0 (see \citeN{Doornik(09)}).
This work was supported by JSPS KAKENHI Grant Numbers 25245035, 26245028, 17H00985, 15H01943, 19H00588.
\bibliography{library1,library2}

\newpage
\setcounter{page}{1}
\begin{center}
\LARGE \bf Supplementary Material
\end{center}
\appendix
\section{Proofs}
%
%
%
%
%
\subsection{Proof of Proposition 5.1}

We first establish the following lemma which describes a property of the local conditional SMC.
%
%
%
%

\begin{lem}
\label{lem 1}
For any $t$ and $t_0$ ($t-K \leq t_0 \leq t$) ,
\begin{eqnarray}
\lefteqn{
\frac{\pi(x_{s-1:t-K-1},x_{t-K}^{k_{t-K}},\ldots,x_{t_0}^{k_{t_0}},\theta \mid y_{s-1:t_0}) }{M^{t_0 - (t-K) +1}}
} && \nonumber\\
&&
\hspace{-5mm}
\times
\prod^M_{\substack{m=1\\ m\neq k_{t-K}}}q_{t-K,\theta}(x^m_{t-K} \mid x_{t-K-1},y_{t-K})
\times
\prod^{t_0}_{j=t-K+1}\prod^M_{\substack{m=1 \\ m\neq k_j}}V_{j-1,\theta}^{a^m_{j-1}}q_{j,\theta}(x_j^m\mid x_{j-1}^{a^m_{j-1}},y_{j})
\label{RHS of lem 1}\\
&=&
\pi(x_{s-1:t-K-1}, \theta \mid y_{s-1:t-K-1})\times
\prod^M_{m=1}q_{t-K,\theta}(x^m_{t-K} \mid x_{t-K-1},y_{t-K})
\nonumber\\
&&
\hspace{-5mm}
\times 
\prod^{t_0}_{j=t-K+1}\prod^M_{m=1}V_{j-1,\theta}^{a^m_{j-1}}q_{j,\theta}(x_j^m\mid x_{j-1}^{a^m_{j-1}},y_{j})
\times
V^{k_{t_0}}_{t_0,\theta}
\times
\prod^{t_0}_{j=t-K}\frac{\hat{p}(y_j \mid x_{t-K-1},y_{s-1:j-1},\theta)}{p(y_j \mid y_{s-1:j-1})}, 
\hspace{10mm}\mbox{}\label{LHS of lem 1}
\end{eqnarray}
where 
\begin{eqnarray}
\hat{p}(y_j \mid x_{t-K-1},y_{s-1:j-1},\theta) &=& \frac{1}{M}\sum^M_{m=1}v_{j,\theta}(x_{j-1}^{a^m_{j-1}},x_{j}^m), \quad j=t-K,\dots,t_0,\label{def of conditional likelihood est}
\end{eqnarray}
with $x^{a^m_{t-K-1}}_{t-K-1} = x_{t-K-1}$ and 
$a^{k_j}_{j-1} = k_{j-1}$, $j = t-K+1,\dots,t_0$. 
\end{lem}
\noindent
The probability density (\ref{RHS of lem 1}) corresponds to the target density $\pi_t^*$ of SMC$^2$ in \shortciteN{Chopin2013} which includes the random particle index. For the particle filtering, the forward block sampling considers the density of $x_{t-K:t-1}^{1:M}$ conditional on $(x_{s-1:t-K-1}, \theta)$, while SMC$^2$ considers that of $x_{1:t}^{1:M}$ conditional on $\theta$. Further, the former updates the importance weight for $(x_{s-1:t}, \theta)$ and the latter updates that for $\theta$  sequentially. \\
%

\noindent
{\it Proof of Lemma \ref{lem 1}.} 
\label{sec:proof_lem 1}
Using Bayes' theorem and 
\begin{eqnarray*}
v_{j,\theta}(x^{a^{m}_{j-1}}_{j-1},x_{j}^{m}) = \frac{f_{\theta}(x^{m}_{j} \mid x^{a^m_{j-1}}_{j-1},y_{j-1})g_{\theta}(y_{j} \mid x_{j}^{m})}{q_{j,\theta}(x_{j}^{m} \mid x_{j-1}^{a^m_{j-1}},y_{j})}, \quad j=1,\ldots,M,
\end{eqnarray*}
the numerator of the first term in (\ref{RHS of lem 1}) is 
\begin{eqnarray}
\lefteqn{\pi(x_{s-1:t-K-1},x_{t-K}^{k_{t-K}},\ldots,x_{t_0}^{k_{t_0}},\theta \mid y_{s-1:t_0})} && \nonumber \\
&=& \frac{\pi(x_{s-1:t-K-1}, \theta\mid y_{s-1:t-K-1})}{p(y_{t-K:t_0} \mid y_{s-1:t-K-1})} \prod^{t_0}_{j=t-K} f_{\theta}(x^{k_j}_j \mid x^{k_{j-1}}_{j-1},y_{j-1})g_\theta(y_j \mid x^{k_j}_j) \nonumber\\
&=& \frac{\pi(x_{s-1:t-K-1}, \theta \mid y_{s-1:t-K-1})}{p(y_{t-K:t_0} \mid y_{s-1:t-K-1})}\prod^{t_0}_{j=t-K} v_{j,\theta}(x_{j-1}^{k_{j-1}},x^{k_j}_{j}) \prod^{t_0}_{j = t-K}q_{j,\theta}(x_{j}^{k_{j}}\mid x^{k_{j-1}}_{j-1},y_{j}). 
\hspace{5mm}\mbox{}
\label{result of Bayes}
\end{eqnarray}
Thus we obtain
\begin{eqnarray*}
(\ref{RHS of lem 1}) & = & 
\frac{\pi(x_{s-1:t-K-1},x_{t-K}^{k_{t-K}},\ldots,x_{t_0}^{k_{t_0}},\theta \mid y_{s-1:t_0})}{M^{t_0 - (t-K) +1}}
\times
\prod^{t_0}_{j=t-K}\prod^M_{\substack{m=1 \\ m\neq k_j}}q_{j,\theta}(x_j^m\mid x_{j-1}^{a^m_{j-1}},y_{j})
\\
&&
\times
\prod^{t_0}_{j=t-K+1}\prod^M_{\substack{m=1\\ m\neq k_j}}V_{j-1,\theta}^{a^m_{j-1}}
\\
&=& \frac{\pi(x_{s-1:t-K-1}, \theta \mid y_{s-1:t-K-1})}{M^{t_0 - (t-K) +1}p(y_{t-K:t_0} \mid y_{s-1:t-K-1})}
\times
\prod^{t_0}_{j=t-K}\prod^M_{m=1}q_{j,\theta}(x_j^m\mid x_{j-1}^{a^m_{j-1}},y_{j}) 
\nonumber \\
&& 
\times 
\prod^{t_0}_{j=t-K+1}v_{j-1,\theta}(x_{j-2}^{k_{j-2}},x^{k_{j-1}}_{j-1})\prod^M_{\substack{m=1\\ m\neq k_j}}V_{j-1,\theta}^{a^m_{j-1}}
\times
v_{t_0,\theta}(x_{{t_0}-1}^{k_{{t_0}-1}},x^{k_{t_0}}_{t_0})
\\
&=&
\frac{\pi(x_{s-1:t-K-1}, \theta \mid y_{s-1:t-K-1})}{\prod^{t_0}_{j=t-K} p(y_{j} \mid y_{s-1:j-1})}
 \times
\prod^{t_0}_{j=t-K}\prod^M_{m=1}q_{j,\theta}(x_j^m\mid x_{j-1}^{a^m_{j-1}},y_{j})
\nonumber \\
&& 
\times
\prod^{t_0}_{j=t-K+1}\prod^M_{m=1}V_{j-1,\theta}^{a^m_{j-1}}
\times
V^{k_{t_0}}_{t_0,\theta}
\times 
\prod^{t_0}_{j=t-K}\hat{p}(y_j \mid x_{t-K-1},y_{s-1:j-1},\theta)
\end{eqnarray*}
and the result follows where we substitute (\ref{result of Bayes}) in the second equality, and used the definition of $\hat{p}(y_j \mid x_{t-K-1},y_{s-1:j-1},\theta)$ in the third equality.
\qed
\\
%
%
%
%
%
%

\noindent
Using Lemma \ref{lem 1}, we obtain Proposition 5.1 as follows.\\

\noindent {\it Proof of Proposition 5.1.}\\ 
\label{sec:proof_prop 2}
By applying Lemma \ref{lem 1} with $t_0 = t-1$ to the first three terms of (44),
we obtain
\begin{eqnarray*}
\lefteqn{\hat{\pi}(x_{s-1:t-K-1},x^{1:M}_{t-K:t},a^{1:M}_{t-K:t-1},k^*_t,\theta \mid y_{s-1:t})}&& \\
&=& \pi(x_{s-1:t-K-1}, \theta \mid y_{s-1:t-K-1})
\times
\prod^M_{m=1}q_{t-K,\theta}(x^m_{t-K} \mid x_{t-K-1},y_{t-K})\\
&& 
\times
\prod^{t-1}_{j=t-K+1}\prod^M_{m=1}V_{j-1,\theta}^{a^m_{j-1}}q_{j,\theta}(x_j^m\mid x_{j-1}^{a^m_{j-1}},y_{j})
\times
V^{k_{t-1}}_{t-1,\theta}   
\times
\prod^{t-1}_{j=t-K}\frac{\hat{p}(y_j \mid x_{t-K-1},y_{s-1:j-1},\theta)}{p(y_j \mid y_{s-1:j-1})}
\\
&& \times \hspace{2mm}q_{t,\theta}(x_t^1 \mid x_{t-1}^{k_{t-1}},y_{t})
\times
\prod^M_{m=2} V_{t-1,\theta}^{a^m_{t-1}}q_{t,\theta}(x_t^m \mid x_{t-1}^{a^m_{t-1}},y_{t})
\times
V^{k^*_t}_{t,\theta}
\times
\frac{\hat{p}(y_t \mid x_{t-K-1},y_{s-1:t-1},\theta)}{p(y_t \mid y_{s-1:t-1})} \nonumber\\
&=&\pi(x_{s-1:t-K-1}, \theta \mid y_{s-1:t-K-1})
\times
\prod^M_{m=1}q_{t-K,\theta}(x^m_{t-K} \mid x_{t-K-1},y_{t-K}) \\
&& \times
\prod^{t}_{j=t-K+1}\prod^M_{m=1}V_{j-1,\theta}^{a^m_{j-1}}q_{j,\theta}(x_j^m,x_{j-1}^{a^m_{j-1}}\mid y_{j})
\times
\prod^{t}_{j=t-K}\frac{\hat{p}(y_j \mid x_{t-K-1},y_{s-1:j-1},\theta)}{p(y_j \mid y_{s-1:j-1})}
\times
V^{k^*_t}_{t,\theta},
\end{eqnarray*}
where we note $a^1_{t-1}=k_{t-1}$ and $k_t=1$. Apply Lemma \ref{lem 1} with $t_0 = t$ and $k_{t_0}=k_t^*$ to the last equation and the result follows.
\qed
%
%
%
%
%
%
\vspace{-3mm}
\subsection{Proof of Proposition 5.2}
\label{sec:proof_prop 3}
We first define the probability density function
\small
\begin{eqnarray*}
\lefteqn{\psi_{\theta,0}(x_{t-K:t}^{1:M},a_{t-K:t-1}^{1:M},k^*_t \mid x_{t-K-1},y_{s-1:t})} &&
\\
& \equiv &
\frac{\pi(x_{t-K:t-1}\mid x_{t-K-1},y_{s-1:t-1},\theta)}{M^K}\\
&&
\hspace{-5mm}
\times
\psi_\theta (x_{t-K}^{-k_{t-K}},\ldots,x_{t-1}^{-k_{t-1}},x^{1:M}_{t},a_{t-K}^{-k_{t-K+1}},\ldots, a_{t-1}^{-k_{t}},k^*_t 
\mid x_{t-K-1:t-1}, a_{t-K}^{k_{t-K+1}},\ldots,a_{t-1}^{k_{t}}, y_{t-K:t}),
\end{eqnarray*}
\normalsize
where $(x_{t-K}^{k_{t-K}},\ldots,x_{t-1}^{k_{t-1}})=x_{t-K:t-1}$ and
\begin{eqnarray*}
\pi(x_{t-K:t-1} \mid x_{t-K-1},y_{s-1:t-1},\theta)=
\frac{\pi(x_{s-1:t-1}, \theta \mid y_{s-1:t-1})}
     {\pi(x_{s-1:t-K-1}, \theta \mid y_{s-1:t-1})}.
\end{eqnarray*}
Noting that 
\begin{eqnarray*}
\lefteqn{\hat{p}(y_t \mid x_{t-K-1},y_{s-1:t-1},\theta)\psi_{\theta,0}(x_{t-K:t}^{1:M},a_{t-K:t-1}^{1:M},k^*_t \mid x_{t-K-1},y_{s-1:t}) }&& \\
&=&\hat{\pi}(x_{s-1:t-K-1},x^{1:M}_{t-K:t},a^{1:M}_{t-K:t-1},k^*_t, \theta \mid y_{s-1:t}) 
\frac{p(y_t \mid y_{s-1:t-1})}{\pi(x_{s-1:t-K-1}, \theta\mid y_{s-1:t-1})},
\end{eqnarray*}
where we used the definition of $\hat{\pi}$ in (44),
\begin{eqnarray*}
\lefteqn{E_{\psi_{\theta,0}}\left[ \hat{p}(y_t \mid x_{t-K-1}, y_{s-1:t-1},\theta) \mid x_{t-K-1}, y_{s-1:t}, \theta \right]} && \\
&=& \int \hat{p}(y_t \mid x_{t-K-1}, y_{s-1:t-1},\theta)\psi_{\theta,0}(x_{t-K:t}^{1:M},a_{t-K:t-1}^{1:M},k^*_t\mid x_{t-K-1},y_{s-1:t})
d x_{t-K:t}^{1:M}d a_{t-K:t-1}^{1:M}d k^*_t \\
&=& \int \hat{\pi}(x_{s-1:t-K-1},x^{1:M}_{t-K:t},a^{1:M}_{t-K:t-1},k^*_t,\theta \mid y_{s-1:t}) d x_{t-K:t}^{1:M}d a_{t-K:t-1}^{1:M}d k^*_t
\frac{p(y_t \mid y_{s-1:t-1})}{\pi(x_{s-1:t-K-1}, \theta\mid y_{s-1:t-1})} \\
&=& \frac{\pi(x_{s-1:t-K-1}, \theta \mid y_{s-1:t})p(y_t \mid y_{s-1:t-1})}{\pi(x_{s-1:t-K-1}, \theta\mid y_{s-1:t-1})} 
\\
&=& p(y_t \mid x_{s-1:t-K-1}, y_{s-1:t-1}, \theta)= p(y_t \mid x_{t-K-1}, y_{s-1:t-1}, \theta).
\end{eqnarray*}
Also it is easy to see 
\begin{eqnarray*}
\lefteqn{E[p(y_t \mid x_{t-K-1}, y_{s-1:t-1}, \theta)\mid y_{s-1:t}]} && \\
& = & \int p(y_t \mid x_{t-K-1}, y_{s-1:t-1}, \theta)\pi(x_{t-K-1},\theta\mid y_{s-1:t-1}) d\theta dx_{t-K-1}
= p(y_t \mid y_{s-1:t-1}).
\end{eqnarray*}
\qed
\subsection{Proof of Proposition 5.3}
\label{sec:proof_prop 5}
We first establish the following lemma as in the proof of Proposition 5.1.

\begin{lem} 
\label{lem in rolling est}
For any $t$, $s_0$ ,and $s$ ($s-1 \leq s_0 \leq s+K-1$), 
\begin{eqnarray}
\lefteqn{
\frac{\pi(x_{s_0}^{k_{s_0}},\ldots,x_{s+K-1}^{k_{s+K-1}},x_{s+K:t},\theta\mid y_{s_0:t})}{M^{(s+K-1)-s_0+1} }
} && \nonumber \\
&&
\times
\prod^M_{\substack{m=1 \\ m\neq k_{s+K-1}}} q_{s+K-1,\theta}(x^m_{s+K-1} \mid x_{s+K},y_{s+K-1})
\times
\prod^{s+K-2}_{j=s_0} \prod^M_{\substack{m=1 \\ m\neq k_j}}V^{a^m_{j+1}}_{j+1,\theta}q_{j,\theta}(x^m_j \mid x^{a^m_{j+1}}_{j+1},y_{j}) \nonumber
\\
& = & 
\pi(x_{s+K:t},\theta \mid y_{s+K:t}) 
\times
\prod^M_{m=1} q_{s+K-1,\theta}(x^m_{s+K-1} \mid x_{s+K},y_{s+K-1})
\nonumber \\
&&
\times
\prod^{s+K-2}_{j=s_0} \prod^M_{m=1}V^{a^m_{j+1}}_{j+1,\theta}q_{j,\theta}(x^m_j \mid x^{a^m_{j+1}}_{j+1},y_{j})
\times V^{k_{s_0}}_{s_0,\theta}
\times 
\prod^{s+K-1}_{j=s_0}\frac{\hat{p}(y_j \mid x_{s+K}, y_{j+1:t},\theta)}{p(y_j \mid y_{j+1:t})}
 \nonumber 
\end{eqnarray} 
with $x_{s+K}^{a_{s+K}^m} = x_{s+K}$ and $a^{k_j}_{j+1} = k_{j+1}\ (s_0\leq j \leq s+K-2)$, where
\begin{eqnarray}
\hat{p}(y_j \mid x_{s+K},y_{j+1:t},\theta) &=& \frac{1}{M}\sum^M_{m=1}v_{j,\theta}(x_j^m,x_{j+1}^{a_{j+1}^m}).
\end{eqnarray}
\end{lem}
\noindent
{\it Proof of Lemma \ref{lem in rolling est}.}\\ 
Since
\begin{eqnarray*}
\lefteqn{
\pi(x_{s_0}^{k_{s_0}},\ldots,x_{s+K-1}^{k_{s+K-1}},x_{s+K:t},\theta\mid y_{s_0:t})} && \\
&=&
\frac{\pi(x_{s+K:t},\theta\mid y_{s+K:t})}{p(y_{s_0:s+K-1}\mid y_{s+K:t})}
\prod_{j=s_0}^{s+K-1}p(x_{j}^{k_{j}}|x_{j+1}^{k_{j+1}},\theta)g_{\theta}(y_j|x_{j}^{k_{j}},x_{j+1}^{k_{j+1}})
\\
&=&
\frac{\pi(x_{s+K:t},\theta\mid y_{s+K:t})}{p(y_{s_0:s+K-1}\mid y_{s+K:t})}
\prod_{j=s_0}^{s+K-1}v_{j,\theta}(x_{j}^{k_{j}}, x_{j+1}^{k_{j+1}})
\times
\prod_{j=s_0}^{s+K-1}q_{j,\theta}(x_{j}^{k_{j}}|x_{j+1}^{k_{j+1}},y_j),
\end{eqnarray*}
we obtain
\begin{eqnarray*}
\lefteqn{
\frac{\pi(x_{s_0}^{k_{s_0}},\ldots,x_{s+K-1}^{k_{s+K-1}},x_{s+K:t}, \theta \mid y_{s_0:t})}{M^{(s+K-1)-s_0+1} }
} && \nonumber \\
&&
\times
\prod^M_{\substack{m=1 \\ m\neq k_{s+K-1}}} q_{s+K-1,\theta}(x^m_{s+K-1} \mid x_{s+K},y_{s+K-1})
\times
\prod^{s+K-2}_{j=s_0} \prod^M_{\substack{m=1 \\ m\neq k_j}}V^{a^m_{j+1}}_{j+1,\theta}q_{j,\theta}(x^m_j \mid x^{a^m_{j+1}}_{j+1},y_{j}) \nonumber
\end{eqnarray*}
\begin{eqnarray*}
& = & 
\frac{\pi(x_{s+K:t},\theta\mid y_{s+K:t})}{M^{(s+K-1)-s_0+1}p(y_{s_0:s+K-1}\mid y_{s+K:t})}
\times
\prod_{j=s_0}^{s+K-1}\prod_{m=1}^Mq_{j,\theta}(x_{j}^{m}|x_{j+1}^{a_{j+1}^m},y_j)
\\
&&
\times
\prod_{j=s_0-1}^{s+K-2}v_{j+1,\theta}(x_{j+1}^{k_{j+1}}, x_{j+2}^{k_{j+2}})
\times
\prod^{s+K-2}_{j=s_0} \prod^M_{\substack{m=1 \\ m\neq k_j}}V^{a^m_{j+1}}_{j+1,\theta}
\\
& = & 
\frac{\pi(x_{s+K:t},\theta\mid y_{s+K:t})}{p(y_{s_0:s+K-1}\mid y_{s+K:t})}
\times
\prod_{j=s_0}^{s+K-1}\prod_{m=1}^Mq_{j,\theta}(x_{j}^{m}|x_{j+1}^{a_{j+1}^m},y_j)
\\
&&
\times
\prod_{j=s_0}^{s+K-1}
\left\{
\frac{1}{M}\sum_{i=1}^Mv_{j,\theta}(x_j^i,x_{j+1}^{a_{j+1}^i})
\right\}
\times
\prod^{s+K-2}_{j=s_0} \prod^M_{\substack{m=1}}V^{a^m_{j+1}}_{j+1,\theta}
\times
V^{k_{s_0}}_{s_0,\theta}
\\
& = & 
\frac{\pi(x_{s+K:t},\theta\mid y_{s+K:t})}{\prod_{j=s_0}^{s+K-1}p(y_j\mid y_{j+1:t})}
\times
\prod_{j=s_0}^{s+K-1}\prod_{m=1}^Mq_{j,\theta}(x_{j}^{m}|x_{j+1}^{a_{j+1}^m},y_j)
\\
&&
\times
\prod^{s+K-2}_{j=s_0} \prod^M_{\substack{m=1}}V^{a^m_{j+1}}_{j+1,\theta}
\times
V^{k_{s_0}}_{s_0,\theta}
\times
\prod_{j=s_0}^{s+K-1} \hat{p}(y_j|x_{s+K},y_{j+1:t},\theta)
\end{eqnarray*}
\qed \\
\vspace{-3mm}
{\it Proof of Proposition 5.3.}\\

By applying Lemma \ref{lem in rolling est} with $s_0 = s-1$ to the first three terms of the target distribution in (49), we have
\begin{eqnarray*}
\lefteqn{\check{\pi}(x_{s-1:s+K-1}^{1:M},x_{s+K:t},a_{s:s+K-1}^{1:M},k_{s-1},k^*_{s},\theta \mid y_{s-1:t})} && \\
&=& \pi(x_{s+K:t},\theta \mid y_{s+K:t})
\times
\prod^M_{m=1} q_{s+K-1,\theta}(x^m_{s+K-1} \mid x_{s+K},y_{s+K-1:t})
\\
&&
\times 
\prod^{s+K-2}_{j=s-1} \prod^M_{m=1} V^{a^m_{j+1}}_{j+1,\theta}q_{j,\theta}(x^m_j \mid x^{a^m_{j+1}}_{j+1},y_{j}) \times V^{k_{s-1}}_{s-1,\theta}\times \prod^{s+K-1}_{j=s-1}\frac{\hat{p}(y_j \mid x_{s+K},y_{j+1:t},\theta)}{p(y_j \mid y_{j+1:t})} 
  \\
&& 
\times V^{k^*_{s}}_{s} 
\times \frac{p(y_{s-1} \mid y_{s:t})}{\hat{p}(y_{s-1} \mid x_{s+K}, y_{s:t},\theta)}\\
&= &  \pi(x_{s+K:t},\theta \mid y_{s+K:t})
\times
\prod^M_{m=1} q_{s+K-1,\theta}(x^m_{s+K-1} \mid x_{s+K},y_{s+K-1:t})
\\
&&
\times 
\prod^{s+K-2}_{j=s}\prod^M_{m=1}  V^{a^m_{j+1}}_{j+1,\theta} q_{j,\theta}(x^m_j \mid x^{a^m_{j+1}}_{j+1},y_{j})\times V^{k^*_{s}}_{s}
\times 
\prod^{s+K-1}_{j=s}\frac{\hat{p}(y_j \mid x_{s+K}, y_{j+1:t},\theta)}{p(y_j \mid y_{j+1:t})} \\
&& 
\times
\prod^M_{m=1}  V^{a^m_{s}}_{s,\theta} q_{s-1,\theta}(x^m_{s-1}\mid x^{a^m_{s}}_{s},y_{s-1})
\times V^{k_{s-1}}_{s-1,\theta}
\end{eqnarray*}
\begin{eqnarray*}
&= &  
\frac{\pi(x_{s}^{k_{s}^*},\ldots,x_{s+K-1}^{k_{s+K-1}^*},x_{s+K:t},\theta \mid y_{s:t})}{M^K}
\times
\prod^M_{\substack{m=1\\ m\neq k_{s+K-1}^*}} q_{s+K-1,\theta}(x^m_{s+K-1} \mid x_{s+K},y_{s+K-1:t})
\\
&&
\times 
\prod^{s+K-2}_{j=s}\prod^M_{\substack{m=1\\ m\neq k_{j}^*}}  V^{a^m_{j+1}}_{j+1,\theta} q_{j,\theta}(x^m_j \mid x^{a^m_{j+1}}_{j+1},y_{j})
\times
\prod^M_{m=1}  V^{a^m_{s}}_{s,\theta} q_{s-1,\theta}(x^m_{s-1}\mid x^{a^m_{s}}_{s},y_{s-1})
\times V^{k_{s-1}}_{s-1,\theta}
\end{eqnarray*}
where we again applied Lemma \ref{lem in rolling est} with $s_0 = s$ and $k_{s-1}=k^*_{s-1}$ in the last equality.
\qed
%
%
%
%
%
%
\subsection{Proof of Proposition 5.4}
\label{sec:proof_prop 6}
{\it Proof of Proposition 5.4.}\\ 
We first define the probability density function
\begin{eqnarray*}
\lefteqn{\bar{\psi}_{\theta,0}(x_{s-1:s+K-1}^{1:M},a_{s:s+K-1}^{1:M},k_{s-1}, k^*_{s} \mid x_{s+K},y_{s-1:s+K-1})} &&
\\
& \equiv &
\frac{\pi(x_{s-1:s+K-1} \mid x_{s+K},y_{s-1:t},\theta)}{M^{K+1}}\\
&\times&\bar{\psi}_\theta (x_{s-1}^{-k_{s-1}},\ldots,x_{s+K-1}^{-k_{s+K-1}},a_{s}^{-k_{s-1}},\ldots,a_{s+K-1}^{-k_{s+K-2}},k^*_{s} \mid x_{s-1:s+K},a_{s-1}^{k_{s-2}},\ldots,a_{s+K-1}^{k_{s+K-2}}, y_{s-1:s+K-1}),
\end{eqnarray*}
and note that
\begin{eqnarray}
\label{eq:suppl_cond}
\pi(x_{s-1:s+K-1} \mid x_{s+K},y_{s-1:t},\theta)
=\frac{\pi(x_{s-1:s+K-1}, x_{s+K:t}, \theta\mid y_{s-1:t})}
     {\pi(x_{s+K:t},\theta  \mid y_{s-1:t})}.
\end{eqnarray}
Since
\begin{eqnarray*}
\lefteqn{\frac{1}{\hat{p}(y_{s-1} \mid x_{s+K}, y_{s:t},\theta)}
\bar{\psi}_{\theta,0}(x_{s-1:s+K-1}^{1:M},a_{s:s+K-1}^{1:M},k_{s-1}, k^*_{s} \mid x_{s+K},y_{s-1:s+K-1})
}&& \\
&=&
\check{\pi}(x_{s-1:s+K-1}^{1:M},x_{s+K:t},a_{s:s+K-1}^{1:M},k_{s-1},k^*_s,\theta \mid y_{s-1:t})\frac{1}{p(y_{s-1} \mid y_{s:t})\pi(x_{s+K:t},\theta\mid y_{s-1:t})},
\end{eqnarray*}
where we used the definition of $\check{\pi}$ in (49), we obtain
\begin{eqnarray*}
\lefteqn{E_{\bar{\psi}_{\theta,0}}\left[ \hat{p}(y_{s-1} \mid x_{s+K}, y_{s:t},\theta)^{-1} \mid x_{s+K}, y_{s-1:t}, \theta \right]} && \\
&=& \int \frac{1}{\hat{p}(y_{s-1} \mid x_{s+K}, y_{s:t},\theta)}
\bar{\psi}_{\theta,0}(x_{s-1:s+K-1}^{1:M},a_{s:s+K-1}^{1:M},k_{s-1}, k^*_{s} \mid x_{s+K},y_{s-1:s+K-1})\\
&& 
\hspace{90mm}
d x_{s-1:s+K-1}^{1:M}d a_{s:s+K-1}^{1:M}dk_{s-1} d k^*_{s} \\
&=& \int 
\check{\pi}(x_{s-1:s+K-1}^{1:M},x_{s+K:t},a_{s:s+K-1}^{1:M},k_{s-1},k^*_s,\theta \mid y_{s-1:t})d x_{s-1:s+K-1}^{1:M}d a_{s:s+K-1}^{1:M}dk_{s-1} d k^*_{s} \\
&& \times 
\frac{1}{p(y_{s-1} \mid y_{s:t})\pi(x_{s+K:t},\theta\mid y_{s-1:t})}
\\
&=& \frac{\pi(x_{s+K:t},\theta \mid y_{s:t})}
         {p(y_{s-1} \mid y_{s:t})\pi(x_{s+K:t},\theta\mid y_{s-1:t})}
=\frac{1}{p(y_{s-1} \mid x_{s+K:t},y_{s:t},\theta)}
=
\frac{1}{p(y_{s-1} \mid x_{s+K}, y_{s:t},\theta)}
\end{eqnarray*}
where we use Proposition 5.3 in the third equality.
Further, 
\begin{eqnarray*}
E[p(y_{s-1} \mid x_{s+K}, y_{s:t},\theta)^{-1}\mid y_{s-1:t}] 
&=& \int \frac{\pi(x_{s+K},\theta \mid y_{s-1:t})}
         {p(y_{s-1} \mid x_{s+K},y_{s:t},\theta)}
          dx_{s+K}d\theta
\\
&=& \int \frac{\pi(x_{s+K},\theta \mid y_{s:t})}
         {p(y_{s-1} \mid y_{s:t})}
          dx_{s+K}d\theta
=p(y_{s-1} \mid y_{s:t})^{-1}.
\end{eqnarray*}
\qed

\subsection*{Proof of Proposition 5.5}
{\it Proof of Proposition 5.5.}\\ 
Consider the joint marginal density of (45):
\begin{eqnarray}
\lefteqn{\hat{\pi}(x_{s-1:t-K-1},x^{1:M}_{t-K:t_0},a^{1:M}_{t-K:t_0-1},
x^{k^*_{t_0+1}}_{t_0+1},\dots,x^{k^*_t}_t,
k^*_{t_0:t},\theta \mid y_{s-1:t})} &&  \nonumber \\
&=& \frac{\pi(x_{s-1:t-K-1},x^{k^*_{t-K}}_{t-K},\dots,x^{k^*_{t}}_t,\theta \mid y_{s-1:t})}
{M^{K+1}}
\times 
\prod^M_{\substack{m=1\\ m \neq k^*_{t-K}}} q_{t-K,\theta}(x^m_{t-K} \mid x_{t-K-1},y_{t-K}) \nonumber \\
&& \quad \times \prod^{t_0}_{j=t-K+1} \prod^M_{\substack{m=1\\ m \neq k^*_j}}V^{a^m_j}_{j-1} q_{j,\theta}(x^m_j \mid x^{a^m_j}_{j-1},y_{j}), 
\end{eqnarray}
for $t_0=t-1,\ldots,t-K+1$, and 
\begin{eqnarray}
\lefteqn{\hat{\pi}(x_{s-1:t-K-1},x^{1:M}_{t-K},
x^{k^*_{t-K+1}}_{t-K+1},\dots,x^{k^*_t}_t,
k^*_{t-K:t},\theta \mid y_{s-1:t})} && \nonumber \\
& = &\frac{\pi(x_{s-1:t-K-1},x^{k^*_{t-K}}_{t-K},\dots,x^{k^*_{t}}_t,\theta \mid y_{s-1:t})}
{M^{K+1}}
\times 
\prod^M_{\substack{m=1\\ m \neq k^*_{t-K}}} q_{t-K,\theta}(x^m_{t-K} \mid x_{t-K-1},y_{t-K}).
\hspace{1cm}\mbox{}
\end{eqnarray}
Then we obtain
\begin{eqnarray}
\lefteqn{\hat{\pi}(k_{t_0}^*\mid x^{1:M}_{s-1:t-K-1},x^{1:M}_{t-K:t_0},a^{1:M}_{t-K:t_0-1},x^{k^*_{t_0+1}}_{t_0+1},\dots,x^{k^*_t}_t,
k^*_{t_0+1:t}, y_{s-1:t},\theta)} &&  \nonumber \\
&\propto &
\hat{\pi}(x^{1:M}_{s-1:t-K-1},x^{1:M}_{t-K:t_0},a^{1:M}_{t-K:t_0-1},
x^{k^*_{t_0+1}}_{t_0+1},\dots,x^{k^*_t}_t,
k^*_{t_0:t},\theta \mid y_{s-1:t})\nonumber \\
&\propto &
\frac{\pi(x_{s-1:t-K-1},x^{k^*_{t-K}}_{t-K},\dots,x^{k^*_{t_0}}_{t_0},\theta \mid y_{s-1:t_0})}{M^{t_0-(t-K)+1}}
\times 
\prod^M_{\substack{m=1\\ m \neq k^*_{t-K}}} q_{t-K,\theta}(x^m_{t-K} \mid x_{t-K-1},y_{t-K}) \nonumber \\
&\times &\prod^{t_0}_{j=t-K+1} \prod^M_{\substack{m=1 \\ m\neq k^*_j}}V^{a^m_j}_{j-1} q_{j,\theta}(x^m_j \mid x^{a^m_j}_{j-1},y_{j})
\times \prod^{t}_{j=t_0+1} 
f_{\theta}(x_j^{k_j^*} \mid x^{k_{j-1}^*}_{j-1},y_{j-1}) 
g_{\theta}(y_j \mid x^{k_{j}^*}_{j}) 
\nonumber \\
&=&
\pi(x_{s-1:t-K-1},\theta \mid y_{s-1:t-K-1})\times
\prod^M_{m=1}q_{t-K,\theta}(x^m_{t-K} \mid x_{t-K-1},y_{t-K})
\nonumber\\
&&
\hspace{-5mm}
\times 
\prod^{t_0}_{j=t-K+1}\prod^M_{m=1}V_{j-1,\theta}^{a^m_{j-1}}q_{j,\theta}(x_j^m\mid x_{j-1}^{a^m_{j-1}},y_{j})
\times
V^{k_{t_0}^*}_{t_0,\theta}
\times
\prod^{t_0}_{j=t-K}\frac{\hat{p}(y_j \mid x_{t-K-1},y_{s-1:j-1},\theta)}{p(y_j \mid y_{s-1:j-1})}, 
\hspace{10mm}\mbox{}
\nonumber\\
&& 
\times \prod^{t}_{j=t_0+1} 
f_{\theta}(x_j^{k_j^*} \mid x^{k_{j-1}^*}_{j-1},y_{j-1}) 
g_{\theta}(y_j \mid x^{k_{j}^*}_{j}) 
\nonumber\\
&\propto & 
V^{k_{t_0}^*}_{t_0,\theta}
\times f_{\theta}(x_{t_0+1}^{k_{t_0+1}^*} \mid x^{k_{t_0}^*}_{t_0},y_{t_0}), 
\end{eqnarray}
where we use Lemma A.1
at the equality.
\qed

\section{Sequential MCMC estimation without rolling the window}
We first give the initializing algorithm which is obtained by skipping the discarding step (Step 2) in the particle rolling algorithm. Next, we describe how to estimate the marginal likelihood.
\subsection{Algorithm}
\begin{description}
\item[](1) At time $j=1$, sample $(x^{n}_1,\theta^n)$ from $\pi(x_1, \theta \mid y_1)$ for $n=1,\ldots,N$.
\begin{enumerate}
\item Sample $\theta^n \sim p(\theta)$, and  $x^{n,m}_1 \sim q_{1,\theta^n}(\cdot \mid y_1)$ for each $m \in \{1,\ldots,M\}$.
\item Sample $k_1 \sim \mathcal{M}(V^{n,1:M}_{1,\theta^n})$ where
\begin{eqnarray}
V^{n,m}_{1,\theta^n} &=& \frac{v_{1,\theta^n}(x^{n,m}_1) }{\sum^M_{i=1}v_{1,\theta^n}(x^{n,i}_1) }, \quad
 v_{1,\theta^n}(x^{n,m}_1) = \frac{\mu_{\theta^n}(x^{n,m}_1)g_{\theta^n}(y_{1} \mid x_{1}^{n,m})}{q_{1,{\theta^n}}(x_{1}^{n,m} \mid y_{1})}.
\end{eqnarray}
\item
Set $x_1^n=x_1^{n,k_1}$ and store $(x^{n}_{1},\theta^n)$ with its importance weight
\begin{eqnarray}
W_{1}^n &\propto& \hat{p}(y_1 \mid \theta^n), \quad
\hat{p}(y_1 \mid \theta^n) = \sum^M_{m=1}v_{1,\theta^n}(x^{n,m}_1). \label{def of est of the initial likelihood}
\end{eqnarray}
\end{enumerate}
\item[](2) At time $j=2,\ldots,L+1$, implement the forward block sampling to generate $x_{1:j}^n$ and $\theta^n$, and compute its importance weight
\begin{eqnarray}
\label{eq:weight_initial}
W^n_{j} &\propto& \hat{p}(y_{j} \mid x_{j-K-1}^n,y_{1:j-1},\theta^n)\times W^n_{j-1},
\\
&&
\label{eq:phat_initial}
\hat{p}(y_{j} \mid x_{j-K-1}^n,y_{1:j-1},\theta^n)=\frac{1}{M}\sum^M_{m=1}v_{j,\theta^n}(x^{n,a^{n,m}_{j-1}}_{j-1},x_{j}^{n,m}).
\end{eqnarray}
For $j<K$, we set $K=j-1$, and all particles of $x_{1:j}^n$ are resampled. 
\end{description}
\noindent
Especially when $j$ is small and the dimension of $x_{1:j}$ is smaller than that of $\theta$, the MCMC update of $\theta$ could lead to unstable estimation results. We may need to modify the MCMC kernel or skip the update in such a case. 

\subsection{Estimation of the marginal likelihood}
\label{sec: Estimation the marginal likelihood}
As a by-product of the proposed algorithms, we can obtain the estimate of the marginal likelihood defined as
\begin{eqnarray}
p(y_{s:t}) &=& \int p(y_{s:t}\mid x_{s:t},\theta)p(x_{s:t} \mid \theta)p(\theta)dx_{s:t}d \theta,
\end{eqnarray}
so that it is used to compute Bayes factors for model comparison. Since it is expressed as
\begin{eqnarray}
p(y_{s:t}) &=& \frac{p(y_t \mid y_{s-1:t-1})}{p(y_{s-1} \mid y_{s:t})}p(y_{s-1:t-1}),
\end{eqnarray}
we obtain the estimate $\hat{p}(y_{s:t})$ recursively by
\begin{eqnarray}
\hat{p}(y_{s:t}) &=& \frac{\hat{p}(y_t \mid y_{s-1:t-1})}{\hat{p}(y_{s-1} \mid y_{s:t})}\hat{p}(y_{s-1:t-1}),
\end{eqnarray}
where
\begin{eqnarray}
\hat{p}(y_t \mid y_{s-1:t-1}) & = & \sum^N_{n=1} W^{n}_{[s-1,t-1]}\hat{p}(y_t \mid x^n_{t-K-1},y_{s-1:t-1},\theta^n), \\
\hat{p}(y_{s-1} \mid y_{s:t}) & = & \sum^N_{n=1} W^n_{[s-1,t]}\hat{p}(y_{s-1} \mid x^n_{s+K},y_{s:t},\theta^n),
\end{eqnarray}
using (28), (29), (34) and (35).
The initial estimate $\hat{p}(y_{1:L+1})$, $L=t-s$ is given by
\begin{eqnarray}
\hat{p}(y_{1:L+1}) &=& \hat{p}(y_1)\prod^{L+1}_{j=2}\hat{p}(y_j \mid y_{1:j-1}), 
\end{eqnarray}
where we use (\ref{def of est of the initial likelihood}), (\ref{eq:weight_initial}) and (\ref{eq:phat_initial}) to obtain
\begin{eqnarray}
\hat{p}(y_1) & = & \sum^N_{n=1} \hat{p}(y_1 \mid \theta^n)\label{ml est at initial period}, \quad
\hat{p}(y_j \mid y_{1:j-1}) = \sum^N_{n=1} W_{j-1}^n\hat{p}(y_j \mid x_{j-K-1}^n,y_{1:j-1}, \theta^n).
\end{eqnarray}

\section{Additional comparison in the RSV model}
We compare the computation time and the ESS of the practical double-block sampler with those of the MCMC and the particle MCMC. 
For the initial sample period (using $y_{1:1988}$), the MCMC sampling is implemented  with 10,000 iteration (2,000 MCMC samples in the burn-in period are discarded). 
Table \ref{table:mcmc_time} shows the computation times\footnote{The total computation time for the MCMC and the particle MCMC to complete the rolling-window estimation is obtained by multiplying the computation time for the initial sample period by 2261. Thus we obtain $1,293 \times 2,261= 2,923,473$ and $3,189 \times 2,261=7,210,329$ respectively.}
 and ESSs\footnote{The ESS is computed as the average of the ESSs during the rolling estimations for our double-block sampler, while that for each parameter is computed as the MCMC sample size (10,000) divided by the inefficiency factor (defined as $1 + 2\sum^\infty_{s=1} \rho_s$, where $\rho_s$ is the MCMC sample autocorrelation at lag $s$).}  for three methods. 
\begin{table}[H]
\centering
\small
\caption{Computation times for the double-block sampler, MCMC and the particle MCMC.}
\label{table:mcmc_time}
\begin{tabular}{lrlr}
 & Time (seconds) & Param. & ESS \\ \hline
Double-block sampler & {\bf 142,709} & - & {\bf 729}\vspace{2mm}
\\ 
MCMC & ${\bf 2,923,473}$ & $\mu$ & {\bf 35}  \\
 &  & $\phi$ & 1764
 \\
 &  & $\sigma^2_\eta$  & 189
\\
 &  & $\xi$ & 3942
 \\
 &  & $\sigma^2_u$ & 636
 \\
 &  & $\rho$  & 257\vspace{2mm}
\\ 
Particle MCMC & {\bf 7,210,329} & $\mu$  & {\bf 40}
 \\
 &  & $\phi$ & 2184 \\
 &  & $\sigma^2_\eta$ &  221
 \\
 &  & $\xi$ & 4878
 \\
 &  & $\sigma^2_u$ & 656
 \\
 &  & $\rho$  & 189
\\
\hline
\end{tabular}
\normalsize
\end{table}
\noindent
The recursive estimation using the standard MCMC or the particle MCMC takes 20-50 times longer than our proposed method. If we take account of the ESS, it would take 400-900 times longer. These results show that the computation time for our proposed method is much smaller compared with recursive estimations using the standard MCMC or the particle MCMC. 
%

\end{document}